\documentclass[preprint,12pt]{elsarticle}

\usepackage{amsmath,amssymb,amsthm,amsfonts}
\usepackage{mathabx}
\usepackage{mathrsfs}
\usepackage{bm}
\usepackage{graphicx}
\usepackage{color,xcolor}
\usepackage{tabularx}
\biboptions{numbers,sort&compress}
\usepackage{epsfig,psfrag}
\usepackage{tikz}
\usepackage{pgflibraryshapes}
\usepackage{graphicx,psfrag}
\usepackage{hyperref}
\usepackage{subfigure}

\makeatletter
\hypersetup{colorlinks=true}
\AtBeginDocument{\@ifpackageloaded{hyperref}
	{\def\@linkcolor{magenta}
		\def\@anchorcolor{black}
		\def\@citecolor{teal}
		\def\@filecolor{cyan}
		\def\@urlcolor{magenta}
		\def\@menucolor{red}
		\def\@pagecolor{cyan}
		\begingroup
		\@makeother\`%
		\@makeother\=%
		\edef\x{%
			\edef\noexpand\x{%
				\endgroup
				\noexpand\toks@{%
					\catcode 96=\noexpand\the\catcode`\noexpand\`\relax
					\catcode 61=\noexpand\the\catcode`\noexpand\=\relax
				}%
			}%
			\noexpand\x
		}%
		\x
		\@makeother\`
		\@makeother\=
	}{}}
\makeatother

\makeatletter \oddsidemargin 0.0cm \evensidemargin 0.0cm
\newcommand{\be}{\begin{equation}}
\newcommand{\en}{\end{equation}}
\newcommand{\la}{\label}

\newcommand{\paa}{\partial}
\def\rr#1{(\ref{#1})}
\def\bm#1{\mbox{\boldmath{$#1$}}}
\def\ii{{\rm i}}
\def\dd{{\rm d}}

\numberwithin{equation}{section}
\theoremstyle{plain}

\newtheorem{theorem*}{Theorem}

\theoremstyle{definition}

\newcommand{\A}{\mathcal{A}}
\newcommand{\B}{\mathcal{B}}
\DeclareMathOperator{\tr}{tr}
\DeclareMathOperator{\Div}{Div}

\journal{Journal of Sound and Vibration}

\begin{document}

\begin{frontmatter}

\title{\textbf{On propagation of waves in pressurized  fiber-reinforced  hyperelastic tubes based on a reduced model}}

\author[mymainaddress]{Xiang Yu\corref{mycorrespondingauthor}}
\cortext[mycorrespondingauthor]{Corresponding author}
\ead{xianyu3-c@my.cityu.edu.hk}

\author[mysecondaryaddress]{Yibin Fu}

\author[mymainaddress]{Hui-Hui Dai}

\address[mymainaddress]{Department of Mathematics, City University of Hong Kong, Tat Chee Avenue, Kowloon Tong, Hong Kong}

\address[mysecondaryaddress]{School of Computing and Mathematics, Keele University, Staffordshire ST5 5BG, UK}

\begin{abstract}
A refined dynamic finite-strain shell theory for incompressible hyperelastic materials was developed by the authors recently. In this paper, we first derive the associated linearized incremental theory, and then use it to investigate wave propagation in a fiber-reinforced hyperelastic tube that is subjected to an axial pre-stretch and internal pressure.  We obtain the dispersion relations for both axisymmetric and non-axisymmetric waves and discuss their accuracy by comparing them with the exact dispersion relations. The bending effect is also examined by comparing the dispersion curves based on the present theory and membrane theory.  It is shown that the present theory is more accurate than the membrane theory in studying wave propagation and the bending effect plays an important role in some wave modes for relatively large wavenumbers. The effects of the pressure, axial pre-stretch and fiber angle on the dispersion relations are displayed.  These results provide the necessary validation for using the refined dynamic finite-strain shell theory and wave propagation to determine arterial properties.
\end{abstract}

\begin{keyword}
Fiber-reinforced tubes; Hyperelasticity; Dispersion relation; Artery.
\end{keyword}

\end{frontmatter}

\section{Introduction}

Soft biological tissues have attracted much research interest owing to their unusual mechanical properties. Significant contributions have been made to develop mechanically-based hyperelastic models that are capable of accurately describing the behavior of different tissues; see for instance the book by Payan and Ohayon \cite{Payan}.  Among them, arteries are of great concern. The arterial wall consists of three main components: elastin and collagen fibers, gelatinous ground substance and smooth muscle cells. Normally, arteries are hydrated because of the ground substance so that they are practically incompressible. The collagen fibers are initially wavy or crimped. The stiffness of the tissue is modest and  is mainly provided by the elastin fibers. However, when a load is applied and the collagen fibers start to be stretched, the  stiffness of the artery  increases sharply \cite{Ama2018,Ama2019}. So, arteries are generally modeled as fiber-reinforced incompressible hyperelastic materials \cite{Hol} with tubular geometry. Also, arteries are subjected to pre-stresses/pre-deformations due to blood pressure. Therefore, in order to evaluate  material properties  or determine pre-stressed states of arteries from wave velocities \cite{Mak,LC}, it is desirable to investigate wave propagation in pressurized fiber-reinforced incompressible hyperelastic tubes.

The study of wave propagation in elastic tubes dates back to 1876 with the work of Pochhammer \cite{Poch} and Chree
\cite{Chree}, who first investigated the propagation of free harmonic waves in infinitely-long isotropic cylinders, and later Ghosh \cite{Ghosh} extended their treatments to wave propagation in hollow cylinders.  Herrmann and Mirsky \cite{HM} investigated waves in a hollow cylinder under the restriction of axial symmetry of motion. The whole  spectrum solution (i.e., the axisymmetric and non-axisymmetric wave modes) for hollow cylinders was given by Gazis \cite{Gaza} using Helmholtz displacement potentials  and numerically solved at the same time \cite{Gazb}. These works were carried out based on the linear theory of elasticity, and a full treatment of wave propagation in cylindrical structures can be found in the books by Achenbach \cite{Ach} and Auld \cite{Auld}. Recently,  Wu {\it et al.} \cite{Wu} analyzed the axisymmetric torsional waves and longitudinal waves in a pressurized functionally graded elastomeric hollow cylinder by employing the small-on-large theory; the effects of the pressure difference, material gradient, and axial pre-stretch on both the torsional and longitudinal wave propagation characteristics are discussed in detail through numerical examples. Based on Dorfmann and Ogdens's nonlinear framework for electroelasticity and the associated linear incremental theory \cite{Dorf2005,Dorf2006}, Su {\it et al}. \cite{Su} investigated the non-axisymmetric wave propagation in an infinite incompressible soft electroactive hollow cylinder; an exact solution was derived in terms of Bessel functions and numerical examples were presented to show the effects of biasing fields and other parameters on the wave propagation behavior.

Using the Fourier series expansion method, Towfighi {\it et al.} \cite{TKE} solved the dispersion curves of circumferential waves in a transversely isotropic cylindrical pipe, an example of which  is a fiber-reinforced composite cylindrical pipe. In \cite{OS}, Ogden and Singh revisited the problem of wave propagation in an incompressible transversely isotropic solid in the presence of initial stress. Applying a new numerical approach called the state vector and Legendre polynomial hybrid method, Zheng {\it et al.} \cite{ZMLHL} computed the dispersion curves and mode shapes for a general anisotropic hollow cylinder; the effect of the fiber angle on the dispersion characteristics of the hollow cylinder was analyzed in detail in their illustrative examples.

Although the aforementioned works have contributed significantly to wave propagation in elastic tubes, most of them assume that the elastic tube is  isotropic or initially isotropic and  the pre-stressed state is homogeneous. These assumptions are not valid for pressurized arteries since arteries  are anisotropic  due to fiber reinforcement and the pressurized state is inhomogeneous. The investigation of wave propagation in deformed arteries involving both geometric and material nonlinearities  should be based on the small-on-large theory, which assumes that a small time-dependent incremental motion is supposed on a finite deformation \cite{Ogden}.  To the authors' best knowledge, wave propagation in pressurized fiber-reinforced incompressible hyperelastic tubes has not been studied yet in the framework of nonlinear elasticity.

Recently, a refined two-dimensional (2d) dynamic finite-strain shell theory for incompressible hyperelastic materials (abbreviated as refined shell theory in what follows) was developed by the authors  \cite{YFD}. It has the advantages of incorporating both stretching and bending effects and at the same time avoiding complex three-dimensional (3d) computations. We expect that the refined shell theory should be more accurate than the membrane theory, as the latter does not take account of any bending effect. However, it is still  not clear how accurate the refined shell theory is due to the lack of its applications to practical problems.  In this paper, we aim to answer this question by investigating wave propagation in pressurized fiber-reinforced hyperelastic tubes based on the refined shell theory.  The purpose of this paper is two-fold: to quantify the accuracy of the refined shell theory, and  to study wave propagation in pressurized fiber-reinforced hyperelastic tubes.

By linearizing the refined shell theory around a known deformed state, we first derive its linearized incremental theory, which serves as our reduced model for studying wave propagation. Then using the linearized incremental theory, we investigate wave propagation in  a fiber-reinforced incompressible hyperelastic tube subjected to a combined action of axial pre-stretch and internal pressure.  We obtain the dispersion relations for both axisymmetric and non-axisymmetric waves in the tube. By comparing the dispersion curves of axisymmetric waves produced by the present theory and and exact theory, we validate the former and quantify its accuracy.  The bending effect is also examined by comparing the dispersion curves of axisymmetric waves based on the present theory and membrane theory. The results reveal that the present theory is indeed more accurate than the membrane theory in studying wave propagation and the bending effect plays an important role in some wave modes for relatively large wavenumbers. The effects of the pressure, axial pre-stretch and fiber angle on the dispersion relations of axisymmetric and non-axisymmetric waves are exhibited. These established semi-analytic results can be used to determine arterial parameters, like arterial stiffness, from the wave velocity; a similar measurement method has been proposed in  \cite{MCH} as a non-invasive blood pressure measurement technique.

{\bf Notation.} Throughout this paper, boldface letters, for example $\bm{x}$, $\bm{F}$, denote vectors or second-order tensors; calligraphy letters, for example $\mathcal{A}^1$, $\mathcal{A}^2$, represent higher-order tensors. The summation convention for repeated indices is adopted. A comma preceding indices means differentiation and a dot over variables indicates time derivative.  When $\mathcal{A}^1$ is a fourth-order tensor, the notation $\mathcal{A}^1[\bm{F}]$  stands for the second-order tensor with components $\mathcal{A}^1_{ijkl}F_{l k}$; similarly, when $\mathcal{A}^2$ is a sixth-order tensor, the notation $\mathcal{A}^2[\bm{F},\bm{G}]$ stands for the second-order tensor with components $\mathcal{A}^2_{ijkl mn}F_{lk}G_{nm}$.

\section{Summary of a refined shell theory and its application to a benchmark problem}\label{sec:2}

In this section, we present a summary of the refined dynamic finite-strain shell theory  derived in \cite{YFD} and its application to the problem of the extension and inflation of an artery. To facilitate comparison with other studies, the series expansions are carried out around the middle surface   instead of the bottom surface in the original paper, but the steps are essentially the same.

\subsection{A refined two-dimensional dynamic finite-strain shell theory for incompressible hyperelastic materials}
We consider an thin shell of constant thickness $2h$ and composed of incompressible hyperelastic material, which occupies a region $\Omega\times [-h,h]$ in the reference configuration. The thickness $2h$ is assumed to be small when compared with the length scale of  the undeformed middle surface $\Omega$ and its ratio to the radius of curvature is much less than $1$.
The position vectors of a representative particle are denoted by $\bm{X}$ and $\bm{x}$ in the reference and current configurations, respectively.  Following \cite{CG,SDJ},  we parameterize the middle surface $\Omega$ by two curvilinear coordinates $\theta^1$ and $\theta^2$. The position vector of a  point on $\Omega$ is written as $\bm{r}=\bm{r}(\theta^1,\theta^2)$. The (covariant) tangent vectors along the coordinate lines are given by $\bm{g}_\alpha=\partial\bm{r}/\partial\theta^\alpha$, $\alpha=1,2$ whose contravariant counterparts are denoted by $\bm{g}^\alpha$; thus $\bm{g}^\alpha\cdot\bm{g}_\beta=\delta^{\alpha}_\beta$. The unit normal vector $\bm{n}$ to $\Omega$ is defined via the formula $\bm{n}=\bm{g}_1\wedge\bm{g}_2/|\bm{g}_1\wedge\bm{g}_2|$ so that $(\bm{g}_1,\bm{g}_2,\bm{n})$ forms a right-handed basis,  where $\wedge$ means the cross product.

In the reference configuration, the position vector of a material point is decomposed as
\begin{align}
\bm{X}=\bm{r}(\theta^1,\theta^2)+Z\bm{n}(\theta^1,\theta^2),\quad -h\leq Z\leq h,
\end{align}
where $Z$ is the coordinate of the point along the normal direction $\bm{n}$. The curvature map $\bm{\kappa}$ is given by $
\bm{\kappa}=-{\partial\bm{n}}/{\partial\bm{r}}=-\bm{n}_{,\alpha}\otimes\bm{g}^\alpha$, which is symmetric in the sense that $\bm{\kappa}=\bm{\kappa}^T$.
Associated  to $\bm{\kappa}$, the mean and Gaussian curvatures are defined by $H=\frac{1}{2}\tr(\bm{\kappa})$ and $K=\det(\bm{\kappa})$. The map $\bm{\mu}$ appearing in the relation
$\paa \bm{X}/\paa \theta^\alpha= \bm{\mu} \bm{g}_\alpha $ is defined by $\bm{\mu}=\bm{1}-Z\bm{\kappa}$, where $\bm{1}=\bm{I}-\bm{n}\otimes\bm{n}$ denotes the rank-two identity tensor of the tangent plane of $\Omega$, and its determinant is denoted by $\mu(Z)=\det(\bm{\mu})=1-2HZ+KZ^2$. The deformation gradient $\bm{F}=\partial\bm{x}/\partial\bm{X}$ then has the expression
\be \bm{F}= (\nabla\bm{x})\bm{\mu}^{-1}+ \frac{\paa \bm{x}}{\paa Z}\otimes \bm{n}, \la{added1} \en
where $\nabla$ denotes the 2d in-plane gradient on the middle surface $\Omega$ with $\nabla \bm{x}$ given by $\nabla \bm{x}=(\paa \bm{x}/\paa \theta^\alpha) \otimes \bm{g}^\alpha$ and the inverse  of $\bm{\mu}$ can be computed by $\bm{\mu}^{-1}=\bm{1}+Z\bm{\kappa}+Z^2\bm{\kappa}^2+\cdots$.

For the incompressible hyperelastic shell, the equation of motion  and the incompressibility constraint are given by
\begin{align}\label{eq:SFa}
&\Div(\bm{S})+\bm{q}_b=\rho\ddot{\bm{x}},\quad \det(\bm{F})=1  \quad \text{in}\ \Omega\times [-h,h],
\end{align}
where $\bm{q}_b$ is the body force, $\rho$ is the (constant) density,  and $\bm{S}$ is the nominal stress given by
\begin{align}\label{eq:nominal}
\bm{S}=\frac{\partial W}{\partial \bm{F}}-p\bm{F}^{-1},
\end{align}
with $W$ denoting the strain energy function and $p$ the Lagrange multiplier associated with the incompressibility constraint. The traction conditions on the bottom and top surfaces are given by
\begin{align}\label{eq:q-q+}
\bm{S}^T\bm{n}|_{Z=-h}=-\bm{q}^-,\quad \bm{S}^T\bm{n}|_{Z=h}=\bm{q}^+\quad \text{in}\ \Omega,
\end{align}
where  $\bm{q}^-$ and $\bm{q}^+$ are respectively the external loads applied on the bottom and top surfaces.

Assume sufficient smoothness of all quantities involved. The current position vector $\bm{x}=\bm{x}(\bm{r}, Z, t)$ can then be  expanded as a Taylor series around the middle surface $Z=0$:
\begin{align}
\bm{x}=\bm{x}^{(0)}+Z \bm{x}^{(1)}+\frac{1}{2}Z^2 \bm{x}^{(2)}+\cdots,
\end{align}
where here and henceforth the superscript $^{(i)}$ signifies the $i$-th derivative with respect to $Z$ at $Z=0$. Substituting this expansion into the the kinematic relation \rr{added1}
and the constitutive relation \eqref{eq:nominal}, we obtain the following relations among the expansion coefficients
\begin{align}
&\bm{F}^{(0)}=\nabla\bm{x}^{(0)}+\bm{x}^{(1)}\otimes\bm{n},\quad \bm{F}^{(1)}=(\nabla\bm{x}^{(0)})\bm{\kappa}+\nabla\bm{x}^{(1)}+\bm{x}^{(2)}\otimes\bm{n}, \label{eq:Fx}\\
&\bm{S}^{(0)}=W_{\bm{F}}(\bm{F}^{(0)})-p^{(0)}\bm{F}^{(0)-1},\quad \bm{S}^{(1)}=\mathcal{A}^1[\bm{F}^{(1)}]+p^{(0)}\bm{F}^{(0)-1}\bm{F}^{(1)}\bm{F}^{(0)-1}-p^{(1)}\bm{F}^{(0)-1}, \label{eq:SF}
\end{align}
where $W_{\bm{F}}=\partial W/\partial \bm{F}$ and $\mathcal{A}^1=\partial^2 W/\partial \bm{F}^2|_{\bm{F}=\bm{F}^{(0)}}$.

Our reduced model consists of a system of partial differential equations for the components of the leading-order position vector $\bm{x}^{(0)}$ together with the associated boundary conditions.
 These differential equations are derived with the aid of the following four equations:
\begin{align}
\begin{split}\label{eq:A0}
&(1+Kh^2)((W_{\bm{F}}(\nabla\bm{x}^{(0)}+\bm{x}^{(1)}\otimes\bm{n}))^T\bm{n}-p^{(0)}\bm{g})-2Hh^2(\rho\ddot{\bm{x}}^{(0)}-\bm{q}_b^{(0)}-\nabla\cdot\bm{S}^{(0)})\\
&\;\;\;\; -\frac{1}{2}h^2(\bm{\kappa}\bm{g}^\alpha)\cdot\bm{S}^{(0)}_{,\alpha}+\frac{1}{2}h^2(\rho\ddot{\bm{x}}^{(1)}-\bm{q}_b^{(1)}-\nabla\cdot\bm{S}^{(1)})=\bm{m},
\end{split}\\
&\bm{g}\cdot\bm{x}^{(1)}-1=0, \label{eq:R0}
\end{align}
where $\bm{g}=\bm{g}(\bm{x}^{(0)})=\det(\bm{F}^{(0)})\bm{F}^{(0)-T}\bm{n}=\bm{x}^{(0)}_{,1}\wedge\bm{x}^{(0)}_{,2}/|\bm{g}_1\wedge\bm{g}_2|$ and $\bm{m}=(\mu(h)\bm{q}^+-\mu(-h)\bm{q}^-)/2$, and
\begin{align}
&\bm{B}\bm{x}^{(2)}-p^{(1)}\bm{g}+\bm{f}_2=\rho\ddot{\bm{x}}^{(0)},\label{eq:x2}\\
&\bm{g}\cdot\bm{x}^{(2)}+\tr(\bm{F}^{(0)-1}((\nabla\bm{x}^{(0)})\bm{\kappa}+\nabla\bm{x}^{(1)}))=0, \label{eq:p1}
\end{align}
where the matrix $\bm{B}$ and the vector $\bm{f}_2$ are defined by
\begin{align}
\begin{split}\label{eq:B}
&\bm{B}\bm{a}:= (\mathcal{A}^{1}[\bm{a}\otimes \bm{n}]+p^{(0)}\bm{F}^{(0)-1}(\bm{a}\otimes\bm{n})\bm{F}^{(0)-1})^T\bm{n}, \\
& \iff {B}_{ij}=\mathcal{A}^{1}_{3i3j}+p^{(0)}F^{(0)-1}_{3i}F^{(0)-1}_{3j},
\end{split}\\
\begin{split}
&\bm{f}_2 =  (\mathcal{A}^{1}[(\nabla\bm{x}^{(0)})\bm{\kappa}+\nabla\bm{x}^{(1)}]+p^{(0)}\bm{F}^{(0)-1}((\nabla\bm{x}^{(0)})\bm{\kappa}+\nabla\bm{x}^{(1)})\bm{F}^{(0)-1})^T\bm{n}\\
&\qquad+\nabla\cdot\bm{S}^{(0)} +\bm{q}^{(0)}_{b}.
\end{split}
\end{align}
The central idea is to solve \rr{eq:x2} and \rr{eq:p1} to express $(p^{(1)},\bm{x}^{(2)})$ in terms of  $(p^{(0)},\bm{x}^{(1)},\bm{x}^{(0)})$, and to solve \rr{eq:A0} and \rr{eq:R0} to express $(p^{(0)},\bm{x}^{(1)})$ in terms of $\bm{x}^{(0)}$.
Equations \rr{eq:R0} and \rr{eq:p1} are obtained by equating the coefficients of $Z^0$, $Z^1$ in $\eqref{eq:SFa}_2$. To obtain \rr{eq:A0}, we first deduce from \rr{eq:q-q+} the following averaged boundary condition that incorporates the curvature effect:
\begin{align}
\frac{1}{2}(\mu(-h)\bm{S}^T\bm{n}|_{Z=-h}+\mu(h)\bm{S}^T\bm{n}|_{Z=h})=\bm{m}.
\end{align}
Equation \rr{eq:A0} then results from substitution of the expansion for $\bm{S}$ followed by elimination of $ \bm{S}^{(1)T} {\bm n}$ and $\bm{S}^{(2)T} {\bm n}$ from equating the coefficients of $Z^0$ and $Z^1$ in $\eqref{eq:SFa}_1$ (see equations (3.6) and (3.7) in \cite{YFD}). Finally, equation \rr{eq:x2} is obtained by equating the coefficients of $Z^0$ in $\eqref{eq:SFa}_1$ and using \rr{eq:SF}$_2$ to eliminate $\bm{S}^{(1)}$.

We remark  that except for \eqref{eq:A0}, all the other three equations \rr{eq:R0}$-$\rr{eq:p1} are linear in terms of the unknowns to be solved for.
Although \eqref{eq:A0} is nonlinear and cannot be solved analytically in general, we may seek a perturbation solution of the form $p^{(0)}=p^{(0)a}+h^2 p^{(0)b}+\cdots$ and $\bm{x}^{(1)}=\bm{x}^{(1)a}+h^2 \bm{x}^{(1)b}+\cdots$.   By equating the coefficients of $h^0$ in \eqref{eq:A0} and \eqref{eq:R0}, we obtain
\begin{align}
&(W_{\bm{F}}(\nabla\bm{x}^{(0)}+\bm{x}^{(1)a}\otimes\bm{n}))^T\bm{n}-p^{(0)a}\bm{g}=\bm{m},\label{eq:A00}\\
&\bm{g}\cdot\bm{x}^{(1)a}=1.\label{eq:R00}
\end{align}
Once this system of nonlinear algebraic equations is solved for $p^{(0)a}$ and $\bm{x}^{(1)a}$, each higher-order correction can be obtained by solving a system of linear equations.

Finally, by averaging the equation of motion $\eqref{eq:SFa}_1$ followed by the use of the bottom and top traction conditions \rr{eq:q-q+}, one can derive the following equations satisfied by the stress expansion coefficients:
\begin{align}
&\bm{1}\nabla\cdot \bm{S}^{(0)}_t-\bm{\kappa}\bm{S}^{(0)}\bm{n}=(1+\frac{1}{3}Kh^2)\rho  \ddot{\bm{x}}^{(0)}_t -\overline{\bm{q}}_t,\label{eq:ffinal12}\\
\begin{split}
&\nabla \cdot(\bm{1}\bm{S}^{(0)}\bm{n}-\bm{1}\bm{S}^{(0)T}\bm{n})+\tr(\bm{\kappa}\bm{S}^{(0)}_t)+\frac{1}{3}h^2\nabla\cdot(\bm{1}\nabla\cdot\bm{S}^{(1)}_t)\\
=&(1+\frac{1}{3}Kh^2)\rho \ddot{x}^{(0)}_{3} -\overline{q}_{3}+\frac{1}{3}h^2\nabla\cdot(\rho\ddot{\bm{x}}^{(1)}_t-\bm{q}_{bt}^{(1)})-\nabla\cdot\bm{m}_t,\label{eq:ffinal3}
\end{split}
\end{align}
where the subscript $t$ indicates the in-plane part (i.e., $\bm{q}_t=\bm{1}\bm{q}$, $\bm{S}_t=\bm{1}\bm{S}\bm{1})$ and $\overline{\bm{q}}=(\mu(h)\bm{q}^{+}+\mu(-h)\bm{q}^-)/(2h)+(1+Kh^2/3)\bm{q}^{(0)}_b$. These equations correspond to the refined shell equations in \cite{YFD} re-expanded around the middle surface. For readers who are not familiar with the refined shell theory, we provide  an illustrative example of equations \eqref{eq:x2}, \eqref{eq:p1} and \eqref{eq:A00}-\eqref{eq:ffinal3}  specified to a neo-Hookean cylindrical tube in  \ref{app:neo}.  We remark that the term $\frac{1}{3}h^2\nabla\cdot(\bm{1}\nabla\cdot\bm{S}^{(1)}_t)$ in \rr{eq:ffinal3} represents the bending effect, which can be seen from the virtual work principle given in \cite{YFD}. After substitution of the recurrence relations \eqref{eq:A0}-\eqref{eq:p1}, the shell equations \eqref{eq:ffinal12} and \eqref{eq:ffinal3} become a system of differential equations involving $\bm{x}^{(0)}$ only.  Suitable boundary conditions associated with the shell equations \eqref{eq:ffinal12} and \eqref{eq:ffinal3} can be derived based on the 2d shell virtual work principle, which are relegated to \ref{app:bc} since they are not needed in the problem of wave propagation.

\subsection{Application to the extension and inflation of an artery}
Having summarized the refined shell theory, we now demonstrate how it can be applied to obtain asymptotic results for the extension and inflation of an artery. This problem has previously been studied in detail in \cite{YFD}; see also \cite{HO,Hau1979,MH,RM,WF}.

We consider the artery as a thick-walled circular cylindrical tube, which in its undeformed configuration has inner radius $A$ and outer radius $B$; in particular, the radius of the undeformed middle surface is $R_m=(A+B)/2$. When it is uniformly stretched in the axial direction and inflated by an internal pressure $P$, the inner and outer radii become $a$ and $b$, respectively. In terms of cylindrical polar coordinates, the deformation is given by
\begin{align}\label{eq:deformation}
r=r(R),\quad \theta=\Theta,\quad z=\lambda_z X,
\end{align}
where $(R,\Theta,X)$ and $(r,\theta,z)$ are respectively the cylindrical polar coordinates in the undeformed and deformed configurations, and $\lambda_z$ is the uniform stretch in the axial direction. The deformation gradient is  then  given by
\begin{align}\label{eq:FF}
\bm{F}=\frac{r}{R}\bm{e}_\Theta\otimes\bm{e}_{\Theta}+\lambda_z\bm{e}_{X}\otimes\bm{e}_X+r'(R)\bm{e}_R\otimes\bm{e}_{R}.
\end{align}
where $(\bm{e}_R,\bm{e}_\Theta,\bm{e}_X)$ denote the standard basis vectors of cylindrical polar coordinates $(R,\Theta,X)$.

\begin{figure}[h]
	\centering
	\includegraphics[width=0.6\linewidth]{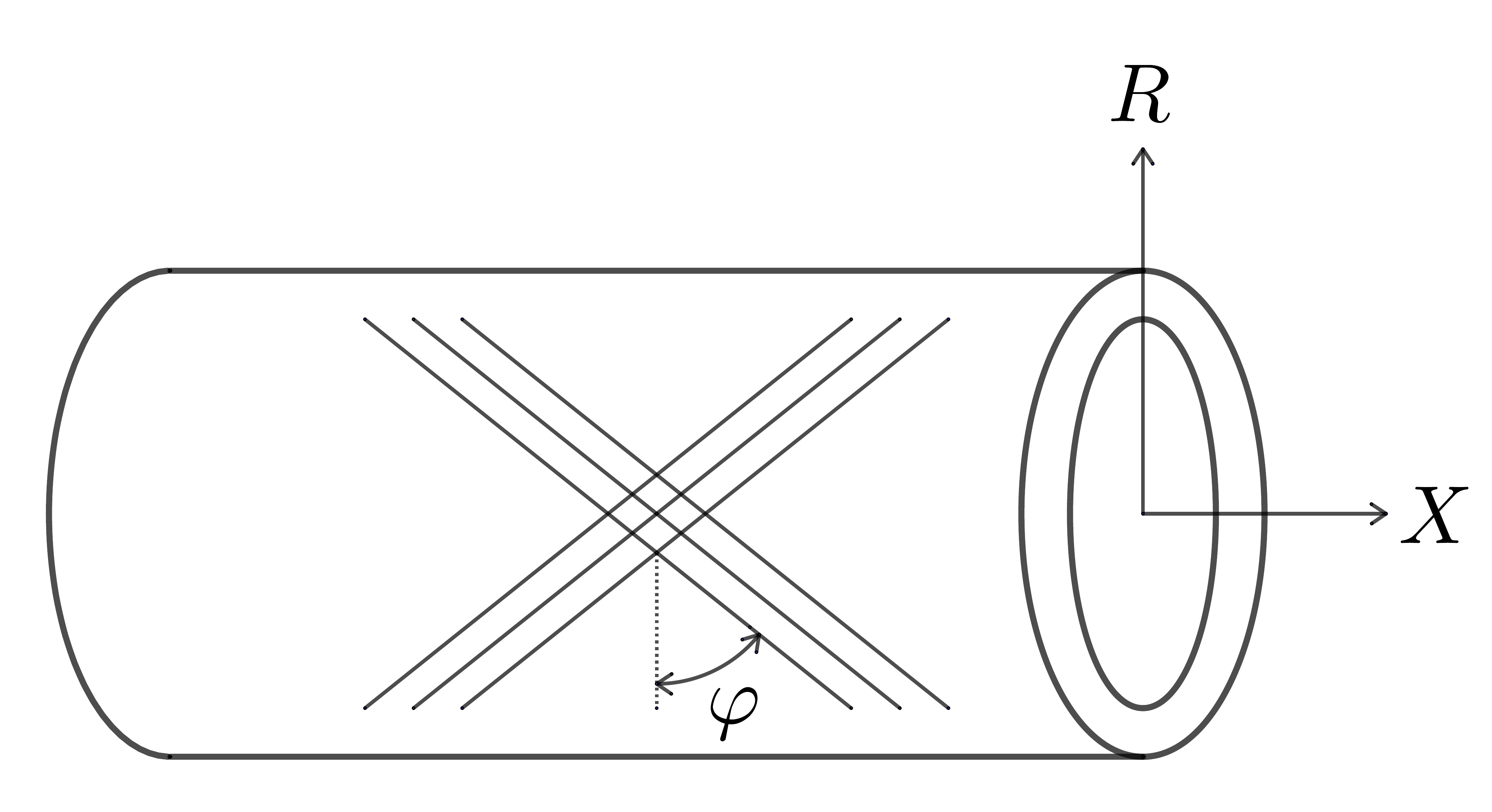}
	\caption{Geometry of a fiber-reinforced cylindrical tube.}
	\label{fig:geometry}
\end{figure}

We assume that the constitutive behavior of the artery is described by a fiber-reinforced incompressible hyperelastic material, for which the strain energy function is given by
\begin{align}\label{eq:SE}
W(I_1,I_4,I_6)=\frac{\mu}{2}(I_1-3)+\frac{k_1}{2k_2}\sum_{i=4,6}(\exp(k_2(I_i-1)^2)-1),
\end{align}
where $\mu$, $k_1$, $k_2$ are material parameters, and the invariants $I_1$, $I_4$ and $I_6$ are defined by
\begin{align}\label{eq:I}
I_1=\tr(\bm{C}),\quad I_4=\bm{M}\cdot (\bm{C}\bm{M}),\quad I_6=\bm{M}'\cdot(\bm{C}\bm{M}').
\end{align}
 In the above expressions $\bm{C}=\bm{F}^T\bm{F}$ is the right Cauchy-Green tensor, and $\bm{M}=\cos\varphi\bm{e}_\Theta+\sin\varphi\bm{e}_X$ and $\bm{M}'=-\cos\varphi\bm{e}_{\Theta}+\sin\varphi\bm{e}_X$ are unit vectors that represent the directions of the two families of fibers which are symmetrically disposed with respect to the axial direction; see Figure \ref{fig:geometry}. The above form of strain energy function is known as the Holzapfel-Gasser-Ogden (HGO) model \cite{Hol}, which is widely used in the constitutive modeling of arterial walls and heart tissues.

The inner surface of the tube is subjected to an internal pressure $P$ and the outer surface is traction-free. Thus we have
\begin{align}
\bm{q}^-=P\bm{F}^{(0)-T}\bm{n}|_{Z=-h}=P\frac{\lambda_za}{A}\bm{e}_R,\quad \bm{q}^+=0.
\end{align}
On the end surface, we impose a resultant axial force
\begin{align}\label{eq:end}
F= 2\pi\int_A^B S_{XX} R\,\dd R-\pi a^2 P.
\end{align}
For the strain energy function \eqref{eq:SE}, the associated nominal stress is given by
\begin{align}\label{eq:SS}
\bm{S}=\mu\bm{F}^T+2k_1(I_4-1)\exp(k_2(I_4-1)^2)(\bm{M}\otimes\bm{F}\bm{M}+\bm{M}'\otimes\bm{F}\bm{M}')-p\bm{F}^{-1}.
\end{align}

To apply the refined shell theory, we expand $r(R)$ into a Taylor series in $Z=R-R_m$
\begin{align}
&r(R)=r_0+r_1Z+\frac{1}{2}r_2 Z^2+\cdots.
\end{align}
Following the procedures detailed in the previous subsection, we obtain the following expressions for $P$ and $F$
\begin{align}
\begin{split}\label{eq:P}
&P=h^*(\mu \frac{\lambda_{m}^4\lambda_{z}^2-1}{ \lambda_{m}^4\lambda_{z}^3}+4\frac{W_4}{\lambda_z}\cos^2\varphi)+h^{*3}\big(\mu \frac{(\lambda_{m}^2\lambda_{z}-1)(\lambda_{m}^2\lambda_{z}+5)}{8\lambda_{m}^8\lambda_{z}^5} \\
&\qquad+\frac{W_4-W_{44}I_1R_m\lambda_{m}^2\lambda_{z}}{2\lambda_{m}^4\lambda_{z}^3}\cos^2\varphi\big),
\end{split}\\
\begin{split}\label{eq:F}
&F=\pi (B^2-A^2)\Big(\mu \frac{2\lambda_m^2\lambda_z^4-\lambda_m^4\lambda_z^2-1}{2\lambda_m^2\lambda_z^3}+2 \frac{W_4}{\lambda_z}(2 \lambda_{z}^2\sin^2\varphi-\lambda_m^2\cos^2\varphi)\\
&\qquad+ h^{*2}\big(\mu \frac{4\lambda_m^2\lambda_z-\lambda_m^4\lambda_z^2-3}{16\lambda_m^6\lambda_z^5}-\frac{W_4
+W_{44}I_1R_m\lambda_m^2\lambda_z}{4\lambda_m^2\lambda_z^3}\cos^2\varphi\big)\Big),
\end{split}
\end{align}
where we have introduced the non-dimensional parameters $h^*=2h/R_m$ and $\lambda_m=r_0/R_m$, and the quantities $I_0$, $I_1$, $W_4$ and $W_{44}$ are given by
\begin{align}
\begin{split}
&I_0=\lambda_{m}^2\cos^2\varphi+\lambda_z^2\sin^2\varphi,\quad I_1=-2(\lambda_{m}^2\lambda_{z}-1)\cos^2\varphi/(\lambda_{z}R_m),\\
&W_4=k_1(I_0-1)\exp(k_2(I_0-1)^2),\quad W_{44}=k_1(1+2k_2(I_0-1)^2)\exp(k_2(I_0-1)^2).
\end{split}
\end{align}

For our numerical calculations later, we adopt the following material parameters: $\mu=74$ (kPa), $k_1=18$ (kPa), $k_2=2.6$, $\varphi=43^\circ$ which are estimated by fitting the biaxial mechanical test data of a human thoracic aorta in \cite{TTS}. We also set $A=8$ (mm) and $B=10$ (mm) which are based on the measured data in \cite{Payan}.

\section{Linearized incremental theory} \label{sec:linearized}
In this section, we derive the incremental equations associated with the refined shell theory  by linearizing the governing equations around a static finitely deformed state described by $\bm{x}=\bm{x}(\bm{X})$ (also called the {\it underlying state}). Suppose that a time-dependent infinitesimal incremental deformation $\delta\bm{x}(\bm{X},t)$ is superposed on this underlying state, where here and henceforth variables with preposed $\delta$ indicate incremental quantities while variables without preposed $\delta$ indicate quantities of the underlying state. First, it follows from \eqref{eq:Fx} and \eqref{eq:SF} that
\begin{align}
\delta\bm{F}^{(0)}&=\nabla \delta\bm{x}^{(0)}+\delta\bm{x}^{(1)}\otimes\bm{n},\quad \delta\bm{F}^{(1)}=(\nabla\delta\bm{x}^{(0)})\bm{\kappa}+\nabla\delta\bm{x}^{(1)}+\delta\bm{x}^{(2)}\otimes\bm{n},\label{eq:iF0F1}\\
\delta\bm{S}^{(0)}&=\mathcal{A}^{1}[\delta\bm{F}^{(0)}]+p^{(0)}\bm{F}^{(0)-1}\delta\bm{F}^{(0)}\bm{F}^{(0)-1}-\delta p^{(0)}\bm{F}^{(0)-1},\label{eq:iS0}\\
\begin{split}\label{eq:iS1}
\delta\bm{S}^{(1)}&=\mathcal{A}^{1}[\delta\bm{F}^{(1)}]+\mathcal{A}^2[\bm{F}^{(1)},\delta\bm{F}^{(0)}] +p^{(1)}\bm{F}^{(0)-1}\delta\bm{F}^{(0)}\bm{F}^{(0)-1} \\
&\quad+ p^{(0)}\bm{F}^{(0)-1}(\delta\bm{F}^{(1)}  -  \bm{F}^{(1)}\bm{F}^{(0)-1}\delta\bm{F}^{(0)}- \delta\bm{F}^{(0)}\bm{F}^{(0)-1}\bm{F}^{(1)})\bm{F}^{(0)-1}\\
&\quad+\delta p^{(0)}\bm{F}^{(0)-1}\bm{F}^{(1)}\bm{F}^{(0)-1} - \delta p^{(1)}\bm{F}^{(0)-1},
\end{split}
\end{align}
where $\mathcal{A}^2=\partial^3 W/\partial \bm{F}^3|_{\bm{F}=\bm{F}^{(0)}}$ denotes the second-order elastic moduli; see \ref{app:moduli} for the expressions of  $\mathcal{A}^1$ and $\mathcal{A}^2$ associated with the anisotropic strain energy function \eqref{eq:SE}. We remark that caution must be taken when dealing with anisotropic materials and one cannot simply adopt the elastic moduli given in \cite{Ogden} (Problem 6.18) which are only valid for isotropic materials.

Next, we turn to find the incremental recurrence relations. Assume that $\delta p^{(0)}=\delta p^{(0)a}+h^2 \delta p^{(0)b}+\cdots$ and $\delta \bm{x}^{(1)}=\delta \bm{x}^{(1)a}+h^2 \delta\bm{x}^{(1)b}+\cdots$. Substituting this asymptotic solution into the linearizations of \eqref{eq:A0} and \eqref{eq:R0} and equating the coefficients of powers of $h$, to leading order, we obtain the following system of linear equations
\begin{align}
&\bm{B}\delta\bm{x}^{(1)a}-\delta p^{(0)a}\bm{g}+\delta \bm{f}_{1}=0,\label{eq:C}\\
&\bm{g}\cdot \delta\bm{x}^{(1)a}+\tr(\bm{F}^{(0)-1}\nabla\delta\bm{x}^{(0)})=0,\label{eq:yx1}
\end{align}
where  $\bm{B}$ is defined as in \eqref{eq:B} and $\delta\bm{f}_1$ is defined by
\begin{align}
\delta\bm{f}_{1}=(\mathcal{A}^{1}[\nabla\delta\bm{x}^{(0)}]+p^{(0)a}\bm{F}^{(0)-1}(\nabla\delta\bm{x}^{(0)})\bm{F}^{(0)-1})^T\bm{n}-\delta\bm{m} \label{eq:f1}
\end{align}
with $\delta\bm{m}=(\mu(h)\delta\bm{q}^+-\mu(-h)\delta\bm{q}^-)/2$. The solution to the above system of linear equations is given by
\begin{align}
&\delta p^{(0)a} = \frac{1}{\bm{g}\cdot\bm{B}^{-1}\bm{g}}(\bm{g}\cdot\bm{B}^{-1}\delta\bm{f}_1-\tr(\bm{F}^{(0)-1}\nabla\delta\bm{x}^{(0)}))\label{eq:ip0},\\
&\delta\bm{x}^{(1)a}=\bm{B}^{-1}(\delta{p}^{(0)a}\bm{g}-\delta\bm{f}_1).\label{eq:ix1}
\end{align}
The next-order solution is a little lengthy and is thus put in \ref{app:corr}. In a similar way, we deduce from the linearizations \eqref{eq:x2} and \eqref{eq:p1} that
\begin{align}
\begin{split}\label{eq:ip1}
\delta p^{(1)} & = \frac{1}{\bm{g}\cdot \bm{B}^{-1}\bm{g}}\big( \bm{g}\cdot \bm{B}^{-1}(\delta\bm{f}_2-\rho\delta\ddot{\bm{x}}^{(0)})\\
&\quad-\tr(\bm{F}^{(0)-1}((\nabla\delta\bm{x}^{(0)})\bm{\kappa}+\nabla\delta\bm{x}^{(1)}-\delta\bm{F}^{(0)}\bm{F}^{(0)-1}\bm{F}^{(1)}))\big),
\end{split}\\
\delta\bm{x}^{(2)} &= \bm{B}^{-1}(\delta p^{(1)}\bm{g}+\rho\delta\ddot{\bm{x}}^{(0)}-\delta\bm{f}_2) \label{eq:ix2}
\end{align}
with  $\delta\bm{f}_2$ given by
\begin{align}
\begin{split}
\delta\bm{f}_2 = &\big(\mathcal{A}^{1}[(\nabla\delta\bm{x}^{(0)})\bm{\kappa}+\nabla\delta\bm{x}^{(1)}]+\mathcal{A}^2[\bm{F}^{(1)},\delta\bm{F}^{(0)}] + p^{(1)}\bm{F}^{(0)-1}\delta\bm{F}^{(0)}\bm{F}^{(0)-1}\\
& +p^{(0)}\bm{F}^{(0)-1}((\nabla\delta\bm{x}^{(0)})\bm{\kappa}+\nabla\delta\bm{x}^{(1)}-\bm{F}^{(1)}\bm{F}^{(0)-1}\delta\bm{F}^{(0)}- \delta\bm{F}^{(0)}\bm{F}^{(0)-1}\bm{F}^{(1)})\bm{F}^{(0)-1}\\
&
+ \delta p^{(0)}\bm{F}^{(0)-1}\bm{F}^{(1)}\bm{F}^{(0)-1}\big)^T\bm{n}+ \nabla\cdot\delta\bm{S}^{(0)}+\delta\bm{q}_b^{(0)}.
\end{split}
\end{align}

Finally, note that the shell equations \eqref{eq:ffinal12} and \eqref{eq:ffinal3} are linear differential equations in terms of the stress expansion coefficients. Thus their linearizations simply take the form
\begin{align}
&\bm{1}\nabla\cdot\delta\bm{S}^{(0)}_t-\bm{\kappa}\delta\bm{S}^{(0)}\bm{n}=(1+\frac{1}{3}Kh^2)\rho\delta\ddot{\bm{x}}^{(0)}_t-\delta\overline{\bm{q}}_t,\label{eq:iffinal12}\\
\begin{split}\label{eq:iffinal3}
&\nabla\cdot(\bm{1}\delta\bm{S}^{(0)}\bm{n}-\bm{1}\delta\bm{S}^{(0)T}\bm{n})+\tr(\bm{\kappa}\delta\bm{S}^{(0)}_t)+\frac{1}{3}h^2\nabla\cdot(\bm{1}\nabla\cdot\delta\bm{S}^{(1)}_t)\\
=&(1+\frac{1}{3}Kh^2)\rho\delta\ddot{x}^{(0)}_3-\delta\overline{q}_3+\frac{1}{3} h^2\nabla\cdot ( \rho \delta\ddot{\bm{x}}^{(1)}_t-\delta\bm{q}_{bt}^{(1)})-\nabla\cdot\delta\bm{m}_{t}
\end{split}\end{align}
with $\delta\overline{\bm{q}}=(\mu(h)\delta\bm{q}^++\mu(-h)\delta\bm{q}^{-})/(2h)+(1+Kh^2/3)\delta\bm{q}_b^{(0)}$, which becomes a system of differential equations in $\delta\bm{x}^{(0)}$ only once  the incremental shell constitutive relations \eqref{eq:iF0F1}-\eqref{eq:iS1} and incremental recurrence relations \eqref{eq:ip0}-\eqref{eq:ix2} are substituted. This completes our derivation of the linearized incremental equations of the refined shell theory.

\section{Propagation of waves in a pressurized and stretched artery}\label{sec:waves}
In this section, we apply the  incremental theory to investigate wave propagation in a pressurized artery with the underlying axisymmetric deformation determined in Section \ref{sec:2}.
As shown in that section, the underlying state is described by $\bm{x}=\lambda_z X\bm{e}_X+r(R)\bm{e}_R$. For convenience, we introduce the notation $u=\delta x^{(0)}_{\Theta}$, $v=\delta x^{(0)}_X$ and $w=\delta x^{(0)}_R$
so that the incremental  vector  $\delta\bm{x}^{(0)}$ can be written as
\begin{align}
\delta\bm{x}^{(0)}=u(X,\Theta,t)\bm{e}_{\Theta}+v(X,\Theta,t)\bm{e}_X+w(X,\Theta,t)\bm{e}_R.
\end{align}
We assume that the traveling wave solution is of the form 
\begin{align}
\begin{split}\label{eq:uvw}
&u(X,\Theta,t)=U\exp(\ii(kX+n\Theta-\omega t)),\\
&v(X,\Theta,t)=V\exp(\ii(kX+n\Theta-\omega t)),\\
&w(X,\Theta,t)=\ii W\exp(\ii(kX+n\Theta-\omega t)),
\end{split}
\end{align}
where $(U,V,W)$ are constants, $k$ and $n$ are the axial and circumferential wavenumbers, respectively, and $\omega$ is the angular frequency. Note that the case $n=0$ corresponds to axisymmetric waves.
In view of the recurrence relations \eqref{eq:ip0}-\eqref{eq:ix2}, the variables $\delta p^{(0)}$, $\delta p^{(1)}$, $\delta \bm{x}^{(1)}$ and $\delta\bm{x}^{(2)}$ can be expressed in terms of $\delta\bm{x}^{(0)}$. Substituting this into the shell constitutive relations \eqref{eq:iS0} and \eqref{eq:iS1}, we obtain expressions of $\delta\bm{S}^{(0)}$ and $\delta\bm{S}^{(1)}$ involving $\delta\bm{x}^{(0)}$ only. The shell equations \eqref{eq:iffinal12} and \eqref{eq:iffinal3} then reduce to a system of linear equations with unknowns $(U,V,W)$:
\begin{align}\label{eq:Mm}
\begin{pmatrix}
m_{11} & m_{12} & m_{13}\\
m_{21} & m_{22} & m_{23}\\
m_{31} & m_{23} & m_{33}
\end{pmatrix}\begin{pmatrix}
U\\
V\\
W
\end{pmatrix}
=\begin{pmatrix}
0\\
0\\
0
\end{pmatrix},
\end{align}
where the coefficients $m_{ij}$ are related to $k$, $n$, $\omega$ and the known quantities of the underlying state, whose  expressions are omitted for  brevity. The existence of a nonzero solution then requires that the determinant of the coefficient matrix $M=(m_{ij})$ must vanish, that is
\begin{align}\label{eq:dispersion}
\det(M)=0,
\end{align}
which gives the desired dispersion relation. This equation  can be solved to express the angular frequency or the phase velocity in terms of the wavenumber and the material parameters
of the artery. In particular, it can be used to determine the blood pressure in an artery by measuring the pressure wave velocity, which was discussed in detail in \cite{MCH} as a non-invasive method for blood pressure measurement. 

The numerical results of the dispersion relations  will be displayed in terms of the following non-dimensional quantities
\begin{align}
k^*=k B,\quad \omega^*=\omega B/\sqrt{\mu/\rho},  \quad c^*=c/\sqrt{\mu/\rho},
\end{align}
where $c=\omega/k$ is the phase velocity. Also, the non-dimensional pressure is defined by $P^*=P/(\mu h^*)$.

\subsection{Exact dispersion relations}

For later validation of the present theory, we provide here a brief description of how to determine the exact dispersion relations of traveling waves, which can be computed by solving an eigenvalue problem based on the linearized theory for incremental deformations superimposed on a finitely deformed configuration. To include non-axisymmetric waves, we consider an incremental position vector of the form
\begin{align}
\delta\bm{x}=u(r,\theta,z,t)\bm{e}_\theta+v(r,\theta,z,t)\bm{e}_z+w(r,\theta,z,t)\bm{e}_r,
\end{align}
where $(r,\theta,z)$ are the cylindrical polar coordinates in the underlying state and 
\begin{align}\label{eq:uvwex}
\begin{split}
&u(r,\theta,z,t)=\hat{u}(r)\exp(\ii(\hat{k}z+n\theta-\omega t)),\\
&v(r,\theta,z,t)=\hat{v}(r)\exp(\ii(\hat{k}z+n\theta-\omega t)),\\
&w(r,\theta,z,t)=\ii\hat{w}(r)\exp(\ii(\hat{k}z+n\theta-\omega t))
\end{split}
\end{align}
with $\hat{k}=k/\lambda_z$. We also assume that the incremental Lagrange multiplier enforcing the incompressibility constraint is given by
\begin{align}\label{eq:inLagrange}
p(r,\theta,z,t)=\ii \hat{p}(r)\exp(\ii(\hat{k}z+n\theta-\omega t)).
\end{align}
Substituting \eqref{eq:uvwex} and \eqref{eq:inLagrange} into the exact linearized incremental equations of motion and boundary conditions, and then simplifying,  we obtain a system of sixth-order linear differential equations and six associated boundary conditions. For our purpose, it is more convenient to write them in matrix form as
\begin{align}
&\frac{\dd\bm{y}}{\dd r}=M(r,n,k,\omega)\bm{y},\quad a\leq r\leq b, \label{eq:bv}\\
&N(r,n,k,\omega)\bm{y}=0,\quad \text{on}\ r=a,b, \label{eq:bv1}
\end{align}
where $\bm{y}=(\hat{u},\hat{u}',\hat{v},\hat{v}',\hat{w},\hat{p})^T$ with the prime denoting $\dd/\dd r$, and the coefficient matrices $M$ and $N$ are given by
\begin{align}\label{eq:exARRA}
M=\begin{pmatrix}
0 & 1 & 0 & 0 & 0 & 0\\
a_{21} & a_{22} & a_{23} & a_{24} & a_{25} & a_{26}\\
0 & 0 & 0 & 1 & 0 & 0\\
a_{41} & a_{41} & a_{43} & a_{44} & a_{45} & a_{46}\\
-n/r & 0 & -k & 0 & -1/r & 0\\
a_{61} & a_{62} & a_{63} & a_{64} & a_{65} & a_{66}
\end{pmatrix},\quad N=\begin{pmatrix}
-1/r & 1 & 0 & 0 & -n/r & 0\\
0 & 0 &  0 & 1 & -k & 0\\
b_{31} & b_{32} & b_{33} & b_{34} & b_{35} & b_{36}
\end{pmatrix}
\end{align}
with the entries $a_{2j}$, $a_{4j}$, $a_{6j}$ and $b_{3j}$, $1\leq j\leq 6$ given in \ref{app:coeff}. When  $n=0$ and $\omega=0$, the eigenvalue problem \eqref{eq:bv} and \eqref{eq:bv1} reduces to its static and axisymmetric counterpart considered in \cite{Hau1979} and \cite{FLF}. The numerical scheme used to solve the current eigenvalue problem is the same as the one adopted in the latter two papers.

\subsection{Dispersion relations of axisymmetric waves}
We first consider the dispersion relations of axisymmetric waves for which $n=0$. In this case, the (incremental) displacements are independent of $\Theta$, and $m_{12}$, $m_{13}$, $m_{21}$ and $m_{31}$ in \eqref{eq:Mm} vanish. In particular, the dispersion equation \eqref{eq:dispersion} can be written as a product of two determinants $D_1D_2=0$ with
\begin{align}
D_1&=m_{11},\quad D_2=\det\begin{pmatrix}
m_{22} & m_{23}\\
m_{32} & m_{33}
\end{pmatrix}. \label{eq:d2}
\end{align}
The equation $D_1=0$ corresponds to the purely circumferential motion in which only the circumferential displacement is non-zero. We call it the {\it circumferential wave}. The equation $D_2=0$ corresponds to motions in which the circumferential displacement is zero while axial and radial displacements are non-zero and are coupled. It will be seen that $D_2=0$ has two branches associated with the {\it axial-radial wave} (lower branch) and {\it radial-axial wave} (upper branch), respectively.
The axial-radial wave refers to the mode that has the asymptotic behavior $W=O(k V)$ as $k\to 0$, whereas the radial-axial wave refers to the mode that has the asymptotic behavior $V=O(k W)$ in the same limit. In \cite{Gaza,Wu} the circumferential wave is referred to as the {\it torsional wave}, and the axial-radial and radial-axial waves are both called the {\it  longitudinal waves}. We also note that the axial-radial wave is referred to as {\it pulse wave} in the medical community, and its speed at $k=0$ can be brought down to zero by finite deformations for some material models, leading to static aneurysm solutions  \cite{FI2015, IF2020}.

The dispersion relations of axisymmetric waves have also been derived based on the membrane assumption. Quoting the results in \cite{FI}, the dispersion equation derived from the membrane theory is
\begin{align}
\begin{split}\label{eq:mem}
&\rho^2\omega^{4} + (\frac{\hat{W}_1-\lambda_1 \hat{W}_{11}}{\lambda_1 R_m^2 }-k^{2}\frac{\lambda_2 \hat{W}_{22}+\hat{W}_2}{\lambda_2 })\rho\omega^{2}+k^{4}\frac{\hat{W}_{22}\hat{W_2}}{\lambda_2 }\\
&+k^{2}\frac{\lambda_2 \hat{W}_{22}(\lambda_1\hat{W}_{11}-\hat{W}_1)-\lambda_1 (\hat{W}_1-\lambda_2 \hat{W}_{12})^2}{\lambda_1\lambda_2^2 R_m^2}=0,
\end{split}
\end{align}
where $\lambda_1=\lambda_m$ and $\lambda_2=\lambda_z$ are the circumferential and axial stretches of the middle surface in the underlying state respectively, $\hat{W}_1=\frac{\partial \hat{W}}{\partial \lambda_1}(\lambda_m,\lambda_z)$, $\hat{W}_2=\frac{\partial \hat{W}}{\partial \lambda_2}(\lambda_m,\lambda_z)$ etc., and $\hat{W}$ is the so-called reduced strain energy function. Corresponding to \rr{eq:SE}, this function takes the form
\begin{align}
\hat{W}(\lambda_1,\lambda_2)=\frac{\mu}{2}(\lambda_1^2+\lambda_2^2+\lambda_1^{-2}\lambda_2^{-2}-3)+\frac{k_1}{k_2}(\exp(k_2(\lambda_1^2\cos^2\varphi+\lambda_2^2\sin^2\varphi-1)^2)-1).
\end{align}
The interested readers are referred to  \cite{EJ} for a very self-contained derivation of the governing equations of membrane theory, which can also be recovered from the present theory by omitting $O(h^2)$-terms in \eqref{eq:iffinal12} and \eqref{eq:iffinal3}.

\subsubsection{Validation of the present theory and the effect of bending stiffness}

Due to the complex expressions of the coefficient matrices $M(r,n,k,\omega,)$ and $N(r,n,k,\omega)$, the exact dispersion relations can only be obtained numerically as stated earlier.  Although the dispersion equation \rr{eq:mem} based on the membrane assumption is analytic, it does not take account of any bending effect. The present theory presents a compromise: it incorporates the bending effect but the associated dispersion relations are still analytic.

\begin{figure}[h]
	\centering
\hspace{-2em}	\subfigure[]{
		\begin{minipage}{4.3cm}
			\centering
			\includegraphics[scale=0.29]{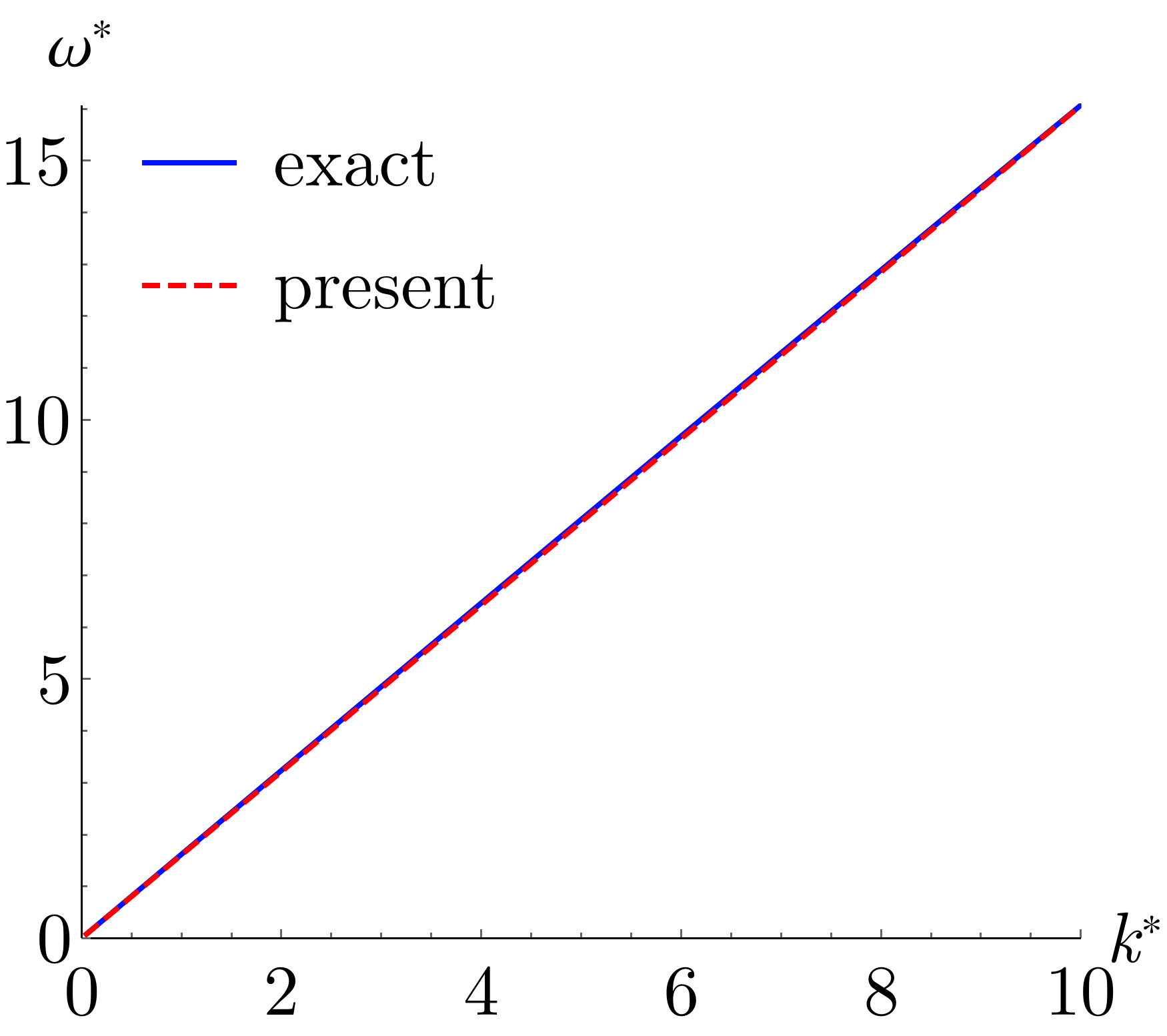}
		\end{minipage}
	}\qquad\quad
	\subfigure[]{
		\begin{minipage}{4.3cm}
			\centering
			\includegraphics[scale=0.29]{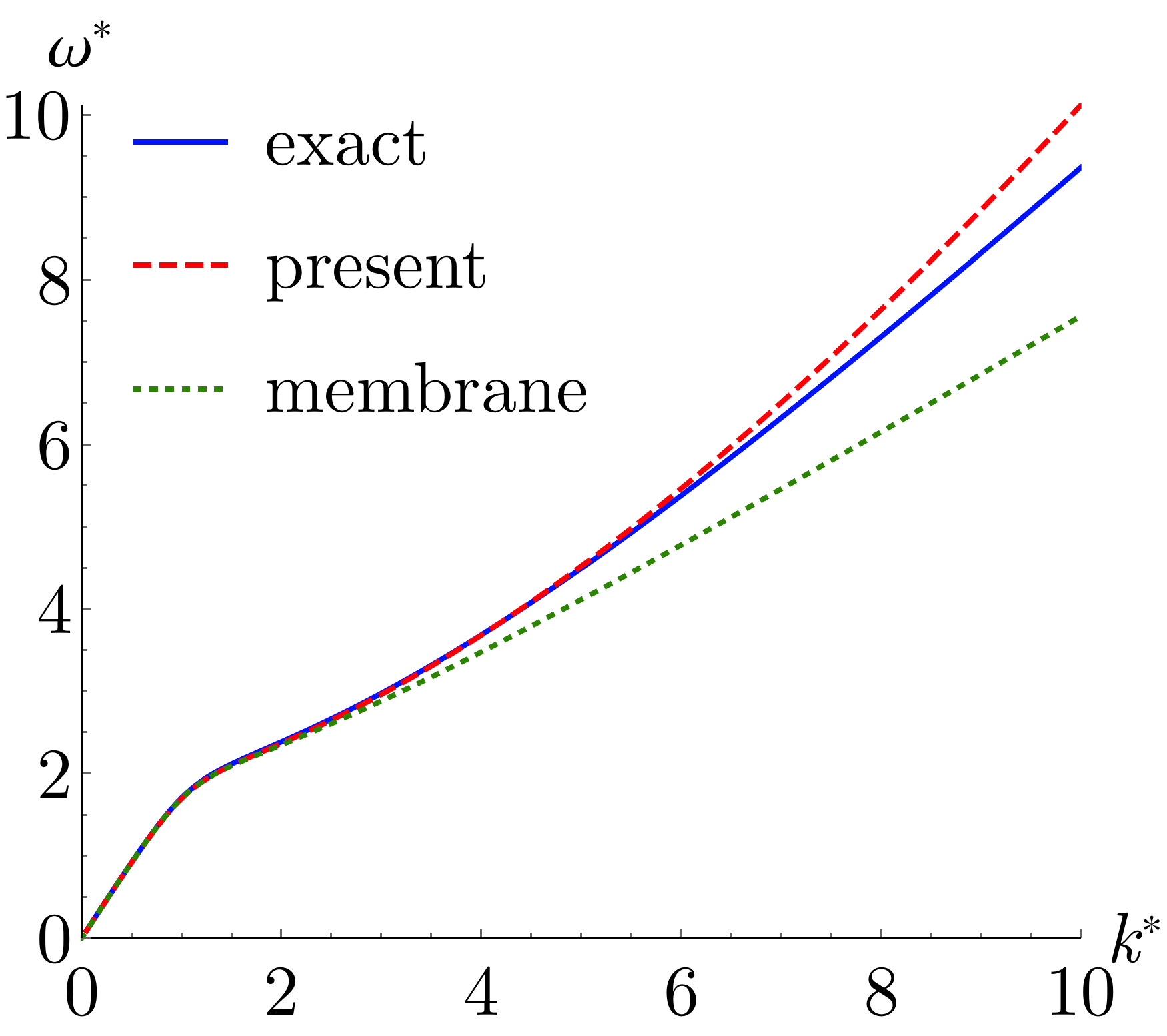}
		\end{minipage}
	}\qquad\quad\subfigure[]{
	\begin{minipage}{4.3cm}
		\centering
		\includegraphics[scale=0.29]{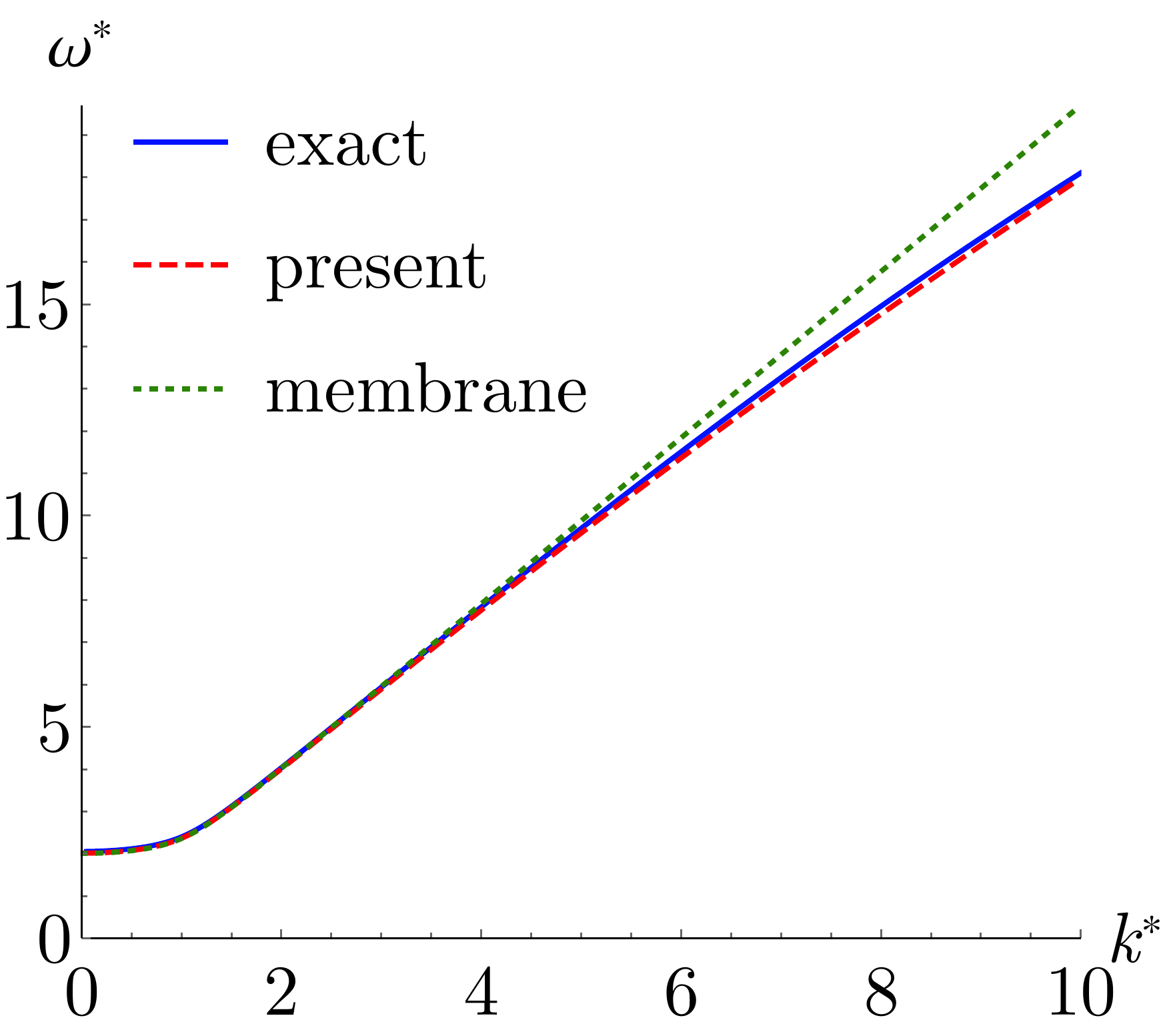}
	\end{minipage}
}
	\caption{(Color online) Comparison of the frequency spectra of axisymmetric waves obtained by the exact theory, present theory and membrane theory: (a) Circumferential wave; (b) Axial-radial wave; (c) Radial-axial wave.}
	\label{fig:e}
\end{figure}

\begin{figure}[h]
	\centering
	\hspace{-3.1em}	\subfigure[]{
		\begin{minipage}{4.3cm}
			\centering
			\includegraphics[scale=0.31]{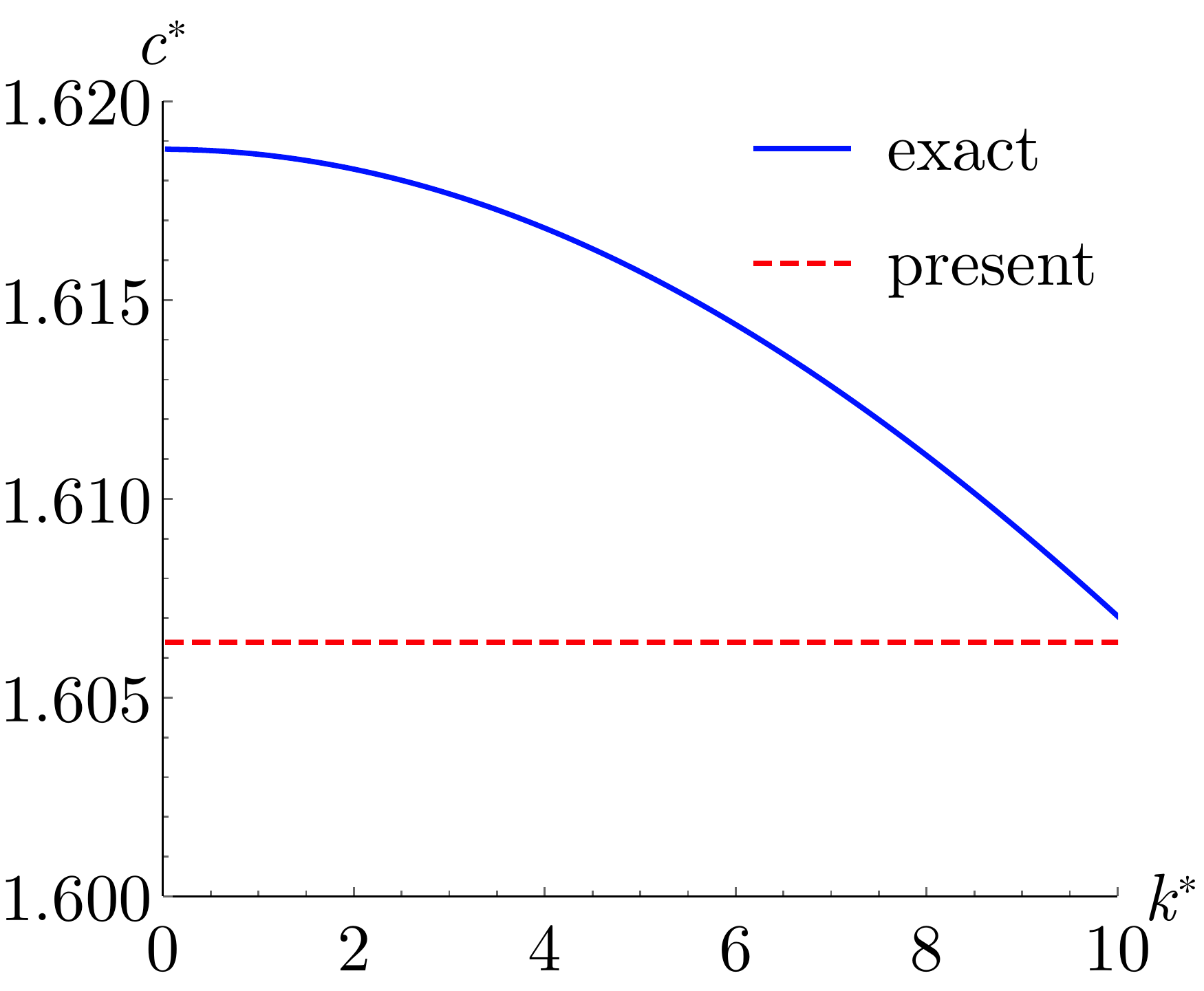}
		\end{minipage}
	}\qquad\quad\hspace{0.6em}
	\subfigure[]{
		\begin{minipage}{4.3cm}
			\centering
			\includegraphics[scale=0.29]{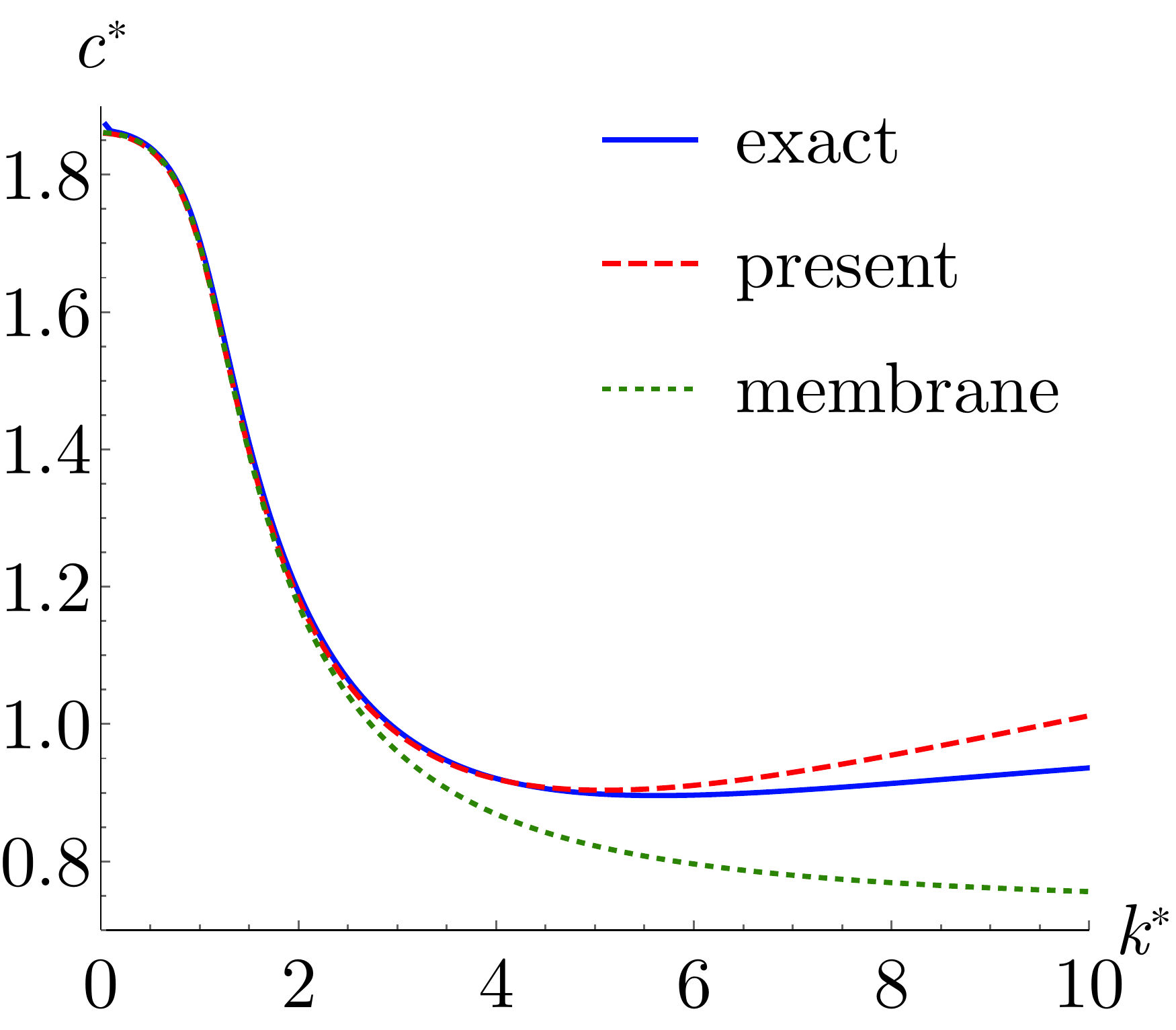}
		\end{minipage}
	}\qquad\quad\subfigure[]{
		\begin{minipage}{4.3cm}
			\centering
			\includegraphics[scale=0.29]{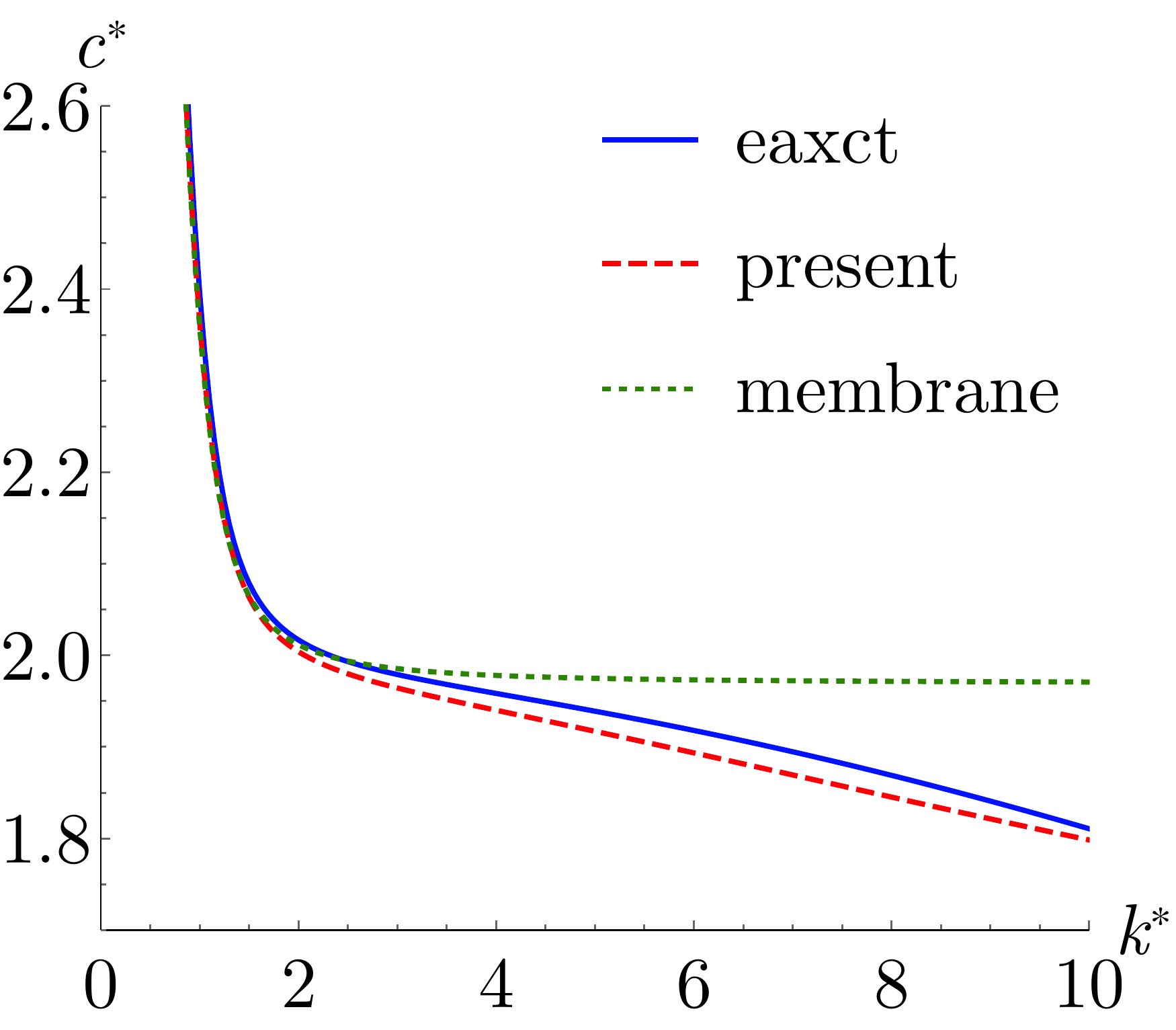}
		\end{minipage}
	}
	\caption{(Color online) Comparison of the phase velocity spectra of axisymmetric waves obtained by the exact theory, present theory and  membrane theory:  (a) Circumferential wave; (b) Axial-radial wave; (c) Radial-axial wave.}
	\label{fig:e1}
\end{figure}

The bending term ($\frac{1}{3}h^2\nabla\cdot(\bm{1}\nabla\cdot\bm{S}_t)$ in \eqref{eq:iffinal3}) ignored in the membrane theory is now singled out in the present 2d shell theory  (in the original 3d theory it is difficult to tell which term represents the bending effect), so by comparing the dispersion relations obtained by the exact theory, present theory   and membrane theory, one can validate the present theory and examine the effect of bending stiffness which may play an important role in some wave modes. For $P^*=0.8$ (corresponding to $P\approx 13$ (kPa), which is a typical value of blood pressure) and $\lambda_z=1$, the frequency spectra and the phase velocities spectra of the circumferential, axial-radial and radial-axial waves in the pressurized artery obtained by the three types of theories are shown in  Figure \ref{fig:e} and Figure \ref{fig:e1}, respectively.  Note that the dispersion curves of the present theory and membrane theory are only meant to be valid in the wavenumber region where $kh\ll 1$ which is equivalent to $ k^*\ll 10$.  Also, since the membrane theory provided in \cite{FI} is not able to capture the circumferential wave mode, there is no result in  Figure \ref{fig:e}(a) or  \ref{fig:e1}(a) corresponding to the membrane theory.
	
 We first compare the dispersion curves based on the present theory  and exact theory. It is seen from Figures \ref{fig:e} and  \ref{fig:e1} that the dispersion curves of the present theory agree very well with the exact dispersion curves in the entire wavenumber range indicated (by which we mean $0\leq k^*\leq 10$) for the circumferential and radial-axial waves  (with relative errors $\leq 0.8\%$ and $\leq 1.9\%$, respectively). For the axial-radial wave, there is excellent agreement between the dispersion curves of present theory and exact theory in the wavenumber range $0\leq k^*\leq 6$ (with a relative error $\leq 1.6\%$) and agreement becomes poor when the wavenumber approaches $10$ (with a relative error $\leq 8.2\%$). In contrast, the membrane theory only provides a good approximation for the exact theory over a much smaller wavenumber region. Thus, to describe wave propagation in arteries where bending effects can be significant due to the moderate thickness of the arterial wall, the present theory may provide an attractive alternative to the exact theory.

As can be seen in the figures, the frequency spectrum of the circumferential wave passes through the origin and are almost straight lines, indicating that this wave mode is almost non-dispersive. On the other hand, the axial-radial and radial-axial waves are dispersive. Furthermore, we can observe that the phase velocity curve of the axial-radial wave starts from a finite value at $k=0$ with zero cut-off frequency, whereas that of the radial-axial wave starts from infinity at $k=0$ with a finite cut-off frequency.  We remark that the bifurcation condition for localized bulging in an inflated elastic tube corresponds to when the phase velocity of the axial-radial wave vanishes  at $k=0$ \cite{FLF}.

\subsubsection{Effect of pressure}
\begin{figure}[h]
	\centering
	\hspace{-2em}	\subfigure[]{
		\begin{minipage}{4.3cm}
			\centering
			\includegraphics[scale=0.29]{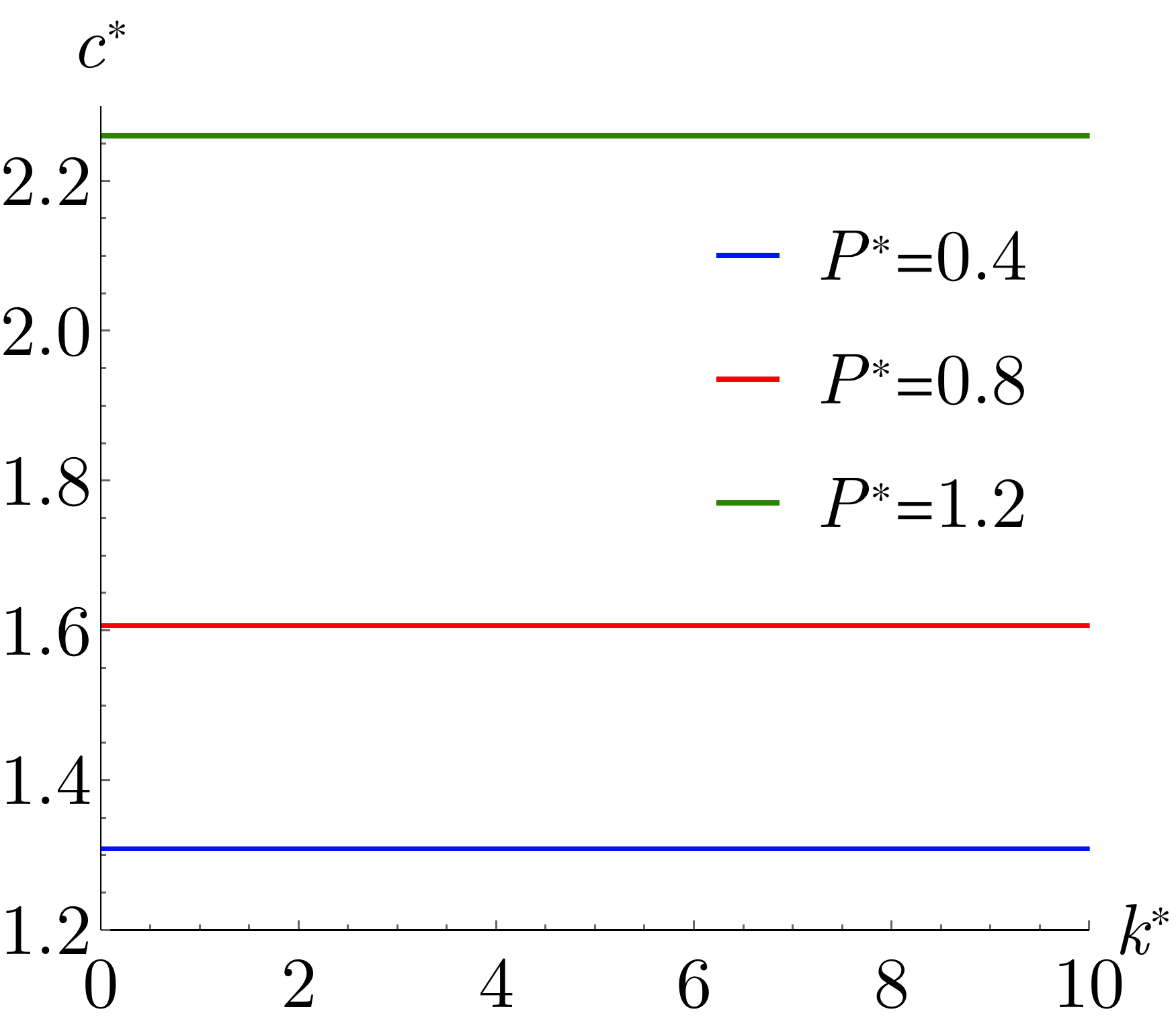}
		\end{minipage}
	}\qquad\quad
	\subfigure[]{
		\begin{minipage}{4.3cm}
			\centering
			\includegraphics[scale=0.29]{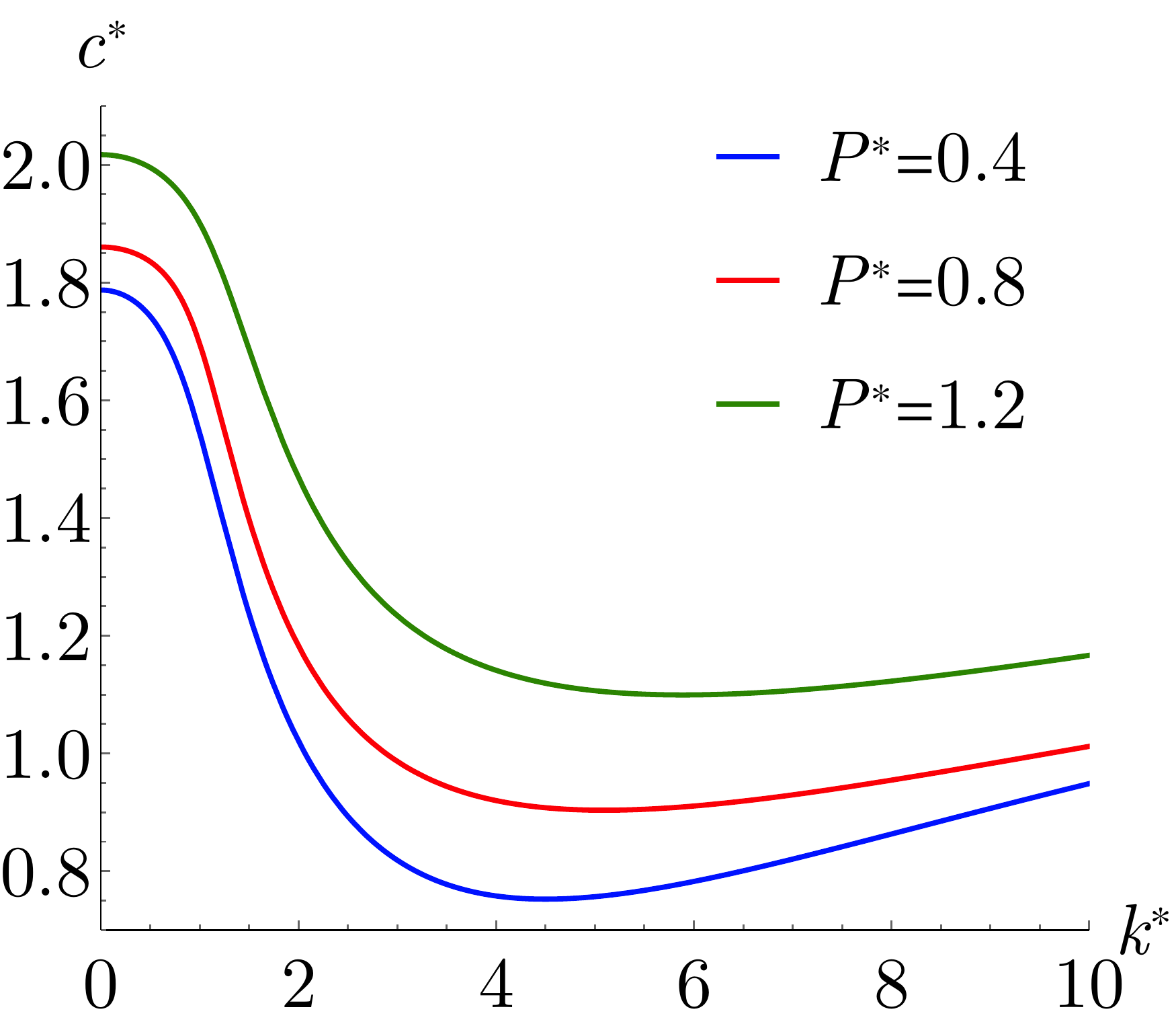}
		\end{minipage}
	}\qquad\quad\subfigure[]{
		\begin{minipage}{4.3cm}
			\centering
			\includegraphics[scale=0.29]{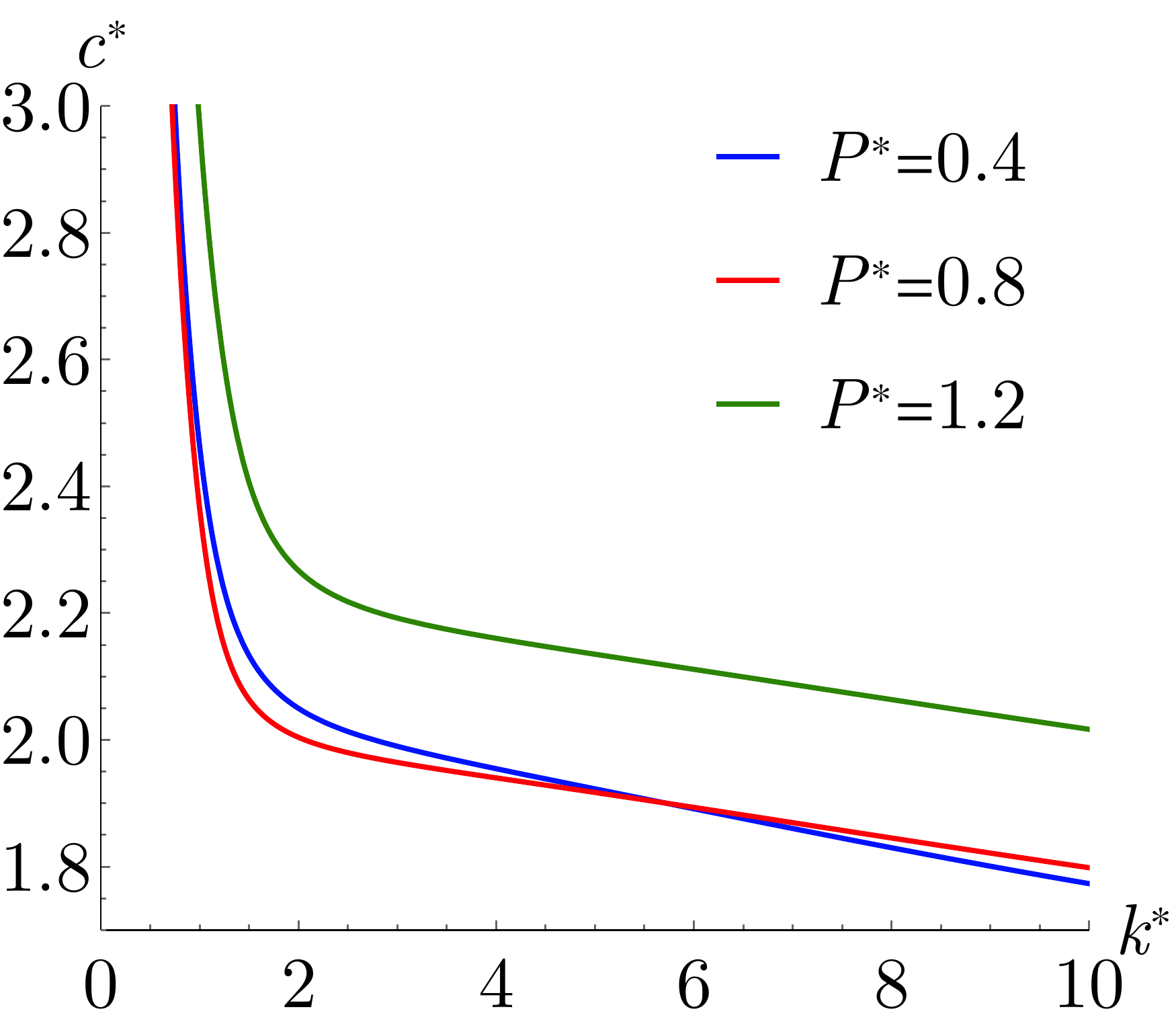}
		\end{minipage}
	}
	\caption{(Color online) Phase velocity spectra of axisymmetric waves at different pressures: (a) Circumferential wave; (b) Axial-radial wave; (c) Radial-axial wave.}
	\label{fig:c1}
\end{figure}

Having validated the present theory, we now use it to obtain the phase velocity spectra for axisymmetric waves propagating in the pressurized artery. For three different values of the pressure $P^*=0.4$, $0.8$ and $1.2$ (corresponding to $P=6.6$ (kPa), $13.1$ (kPa) and $19.7$ (kPa), respectively) with $\lambda_z=1$, the phase velocity spectra of axisymmetric waves are depicted in Figure \ref{fig:c1}. We can observe from the figure that there is an obvious effect of the pressure on the phase velocity spectra of axisymmetric waves. Specifically, the phase velocities of the circumferential and axial-radial waves increase significantly with the pressure in the entire wavenumber range considered. We remark that since the material model under consideration does not allow for aneurysm formation, this increase of velocity of the axial-radial wave with respect to pressure does not contradict our earlier remark that the velocity at $k=0$ may be brought down to zero by finite deformations for material models that admit aneurysm formation. For the radial-axial wave, the increase of the pressure has little impact on its phase velocity when $P^*$ is less than a critical value near $0.8$ and raises its phase velocity noticeably when $P^*$ is greater than the critical value. Additionally,  the phase velocities of the axial-radial wave for all the cases start from a finite value, then arrive at a minimum and subsequently increase gradually with the wavenumber. These non-monotonic phenomena are independent of the pressure and should be attributed to the complex wave interaction with the geometric and material nonlinearities. A similar observation was made by Wu {\it et al.} \cite{Wu} for the axisymmetric wave propagation in a pressurized functionally graded elastomeric hollow cylinder.

\subsubsection{Effects of the axial pre-stretch and fiber angle}
\begin{figure}[h]
	\centering
	\hspace{-2em}	\subfigure[]{
		\begin{minipage}{4.3cm}
			\centering
			\includegraphics[scale=0.29]{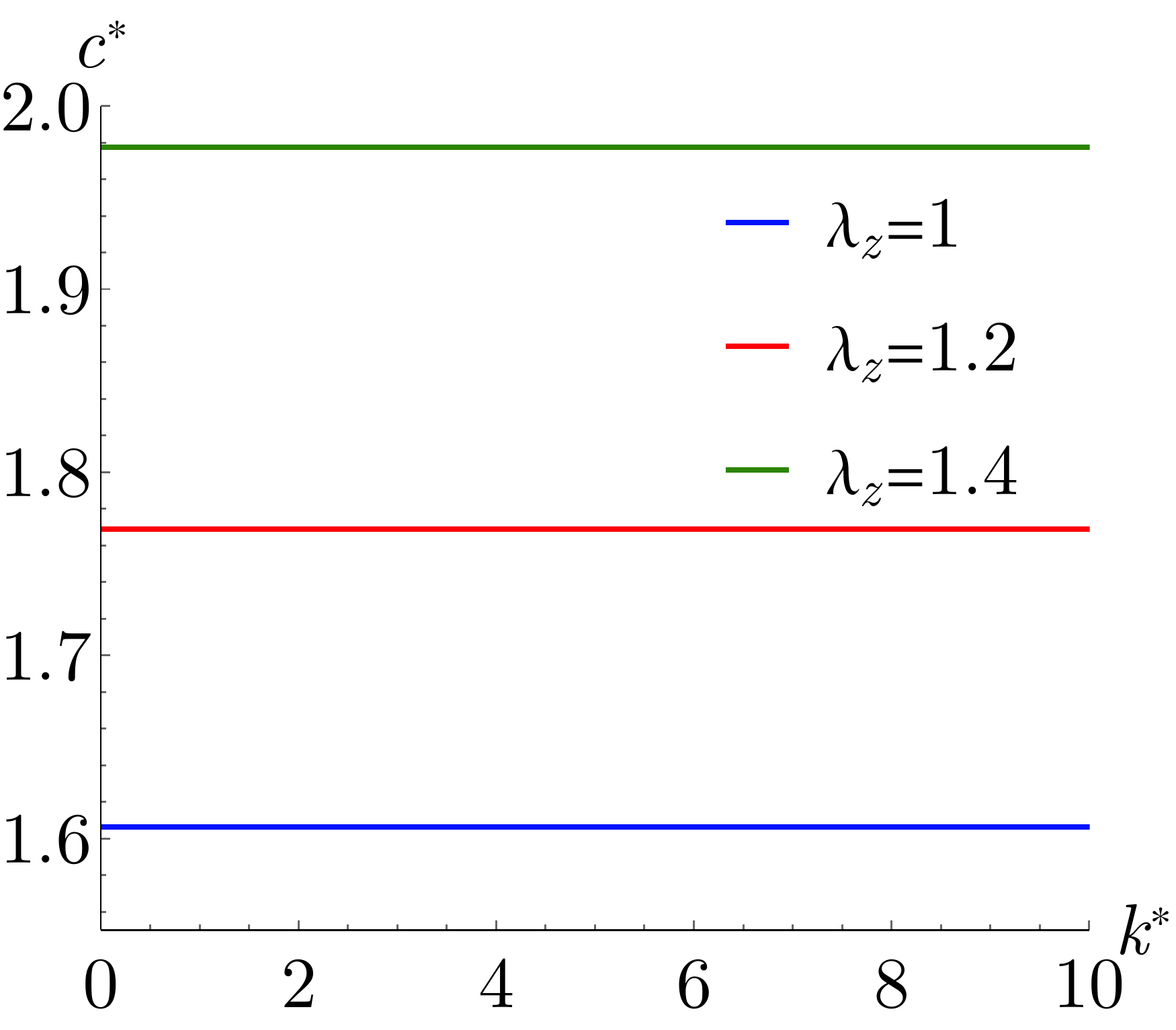}
		\end{minipage}
	}\qquad\quad
	\subfigure[]{
		\begin{minipage}{4.3cm}
			\centering
			\includegraphics[scale=0.29]{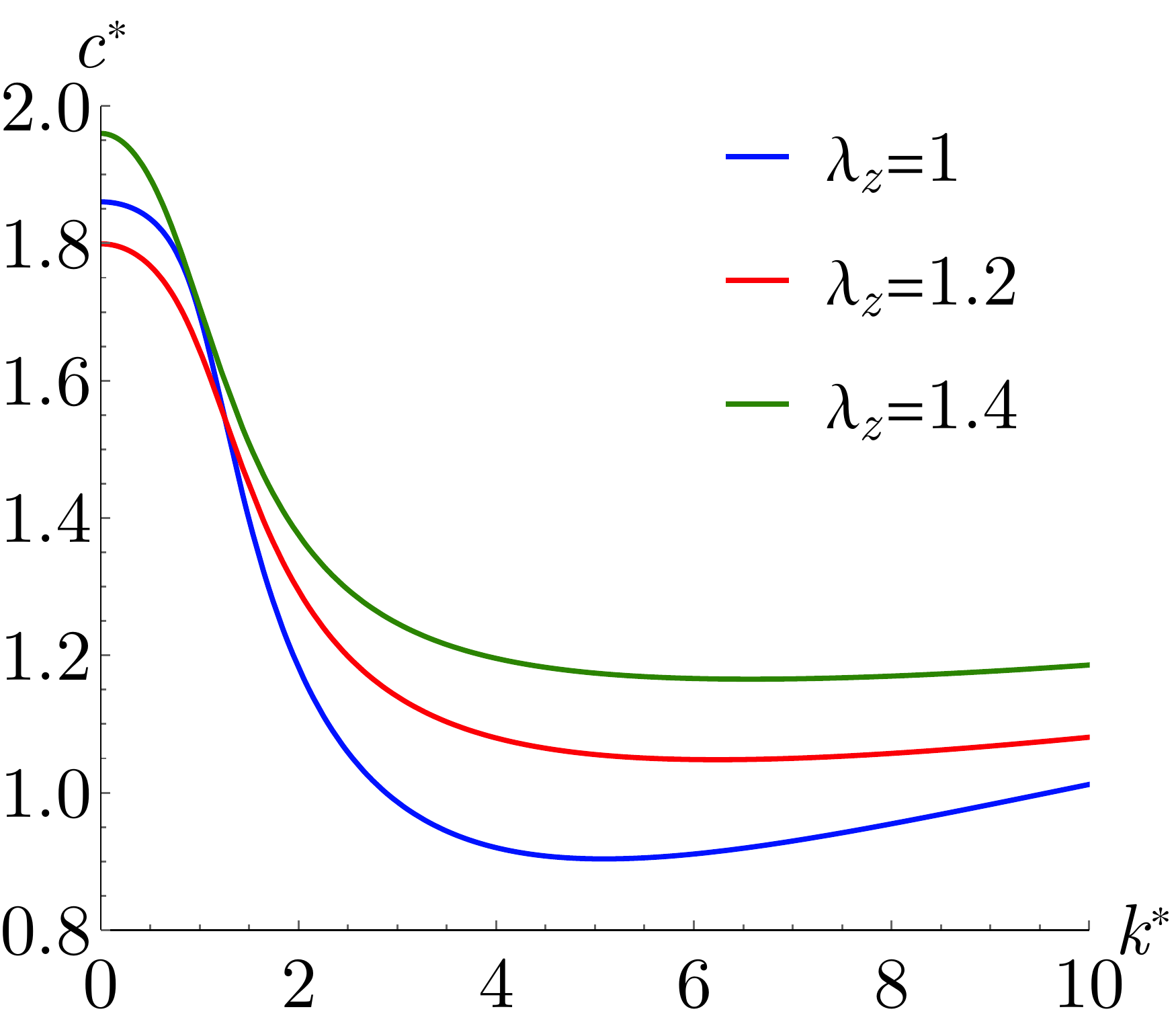}
		\end{minipage}
	}\qquad\quad\subfigure[]{
		\begin{minipage}{4.3cm}
			\centering
			\includegraphics[scale=0.29]{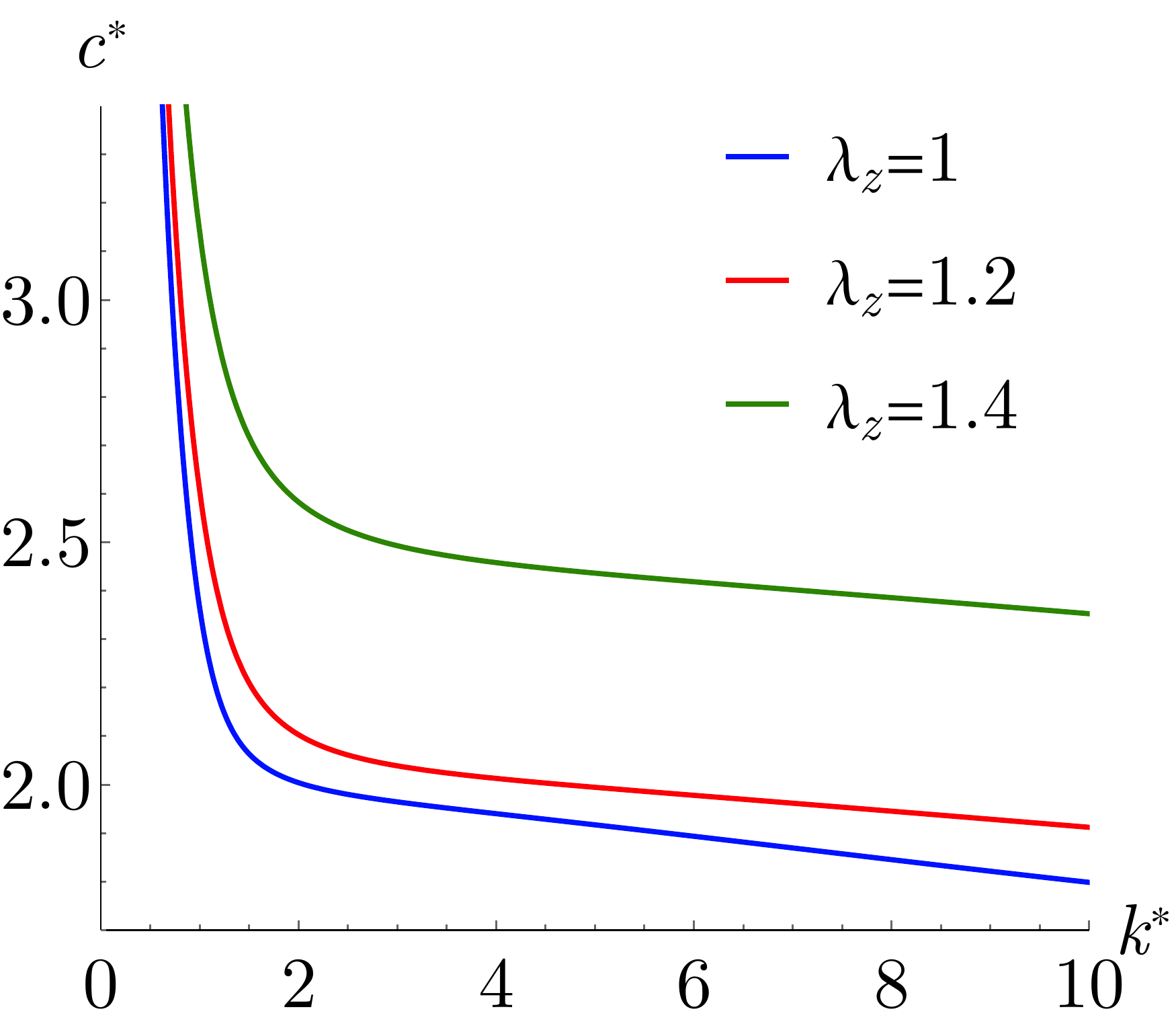}
		\end{minipage}
	}
	\caption{(Color online) Phase velocity spectra of  axisymmetric waves at different axial pre-stretches: (a) Circumferential wave; (b) Axial-radial wave; (c) Radial-axial wave. }
	\label{fig:a1}
\end{figure}

Next, we turn to determine how the axial pre-stretch influences the dispersion relations of axisymmetric waves. For three different values of the axial pre-stretch $\lambda_z=1$, $1.2$ and $1.4$ with  $P^*=0.8$, the phase velocity spectra of axisymmetric waves are presented  in Figure \ref{fig:a1}. As shown in  the figure, the phase velocities of  the circumferential and radial-axial waves increase  with the axial pre-stretch in the entire wavenumber range shown, and that of the axial-radial wave increase with the axial pre-stretch in most of the wavenumber range except in the low wavenumber region, say $k^*\leq 1.5$.

\begin{figure}[h]
	\centering
	\hspace{-2.2em}	\subfigure[]{
		\begin{minipage}{4.3cm}
			\centering
			\includegraphics[scale=0.29]{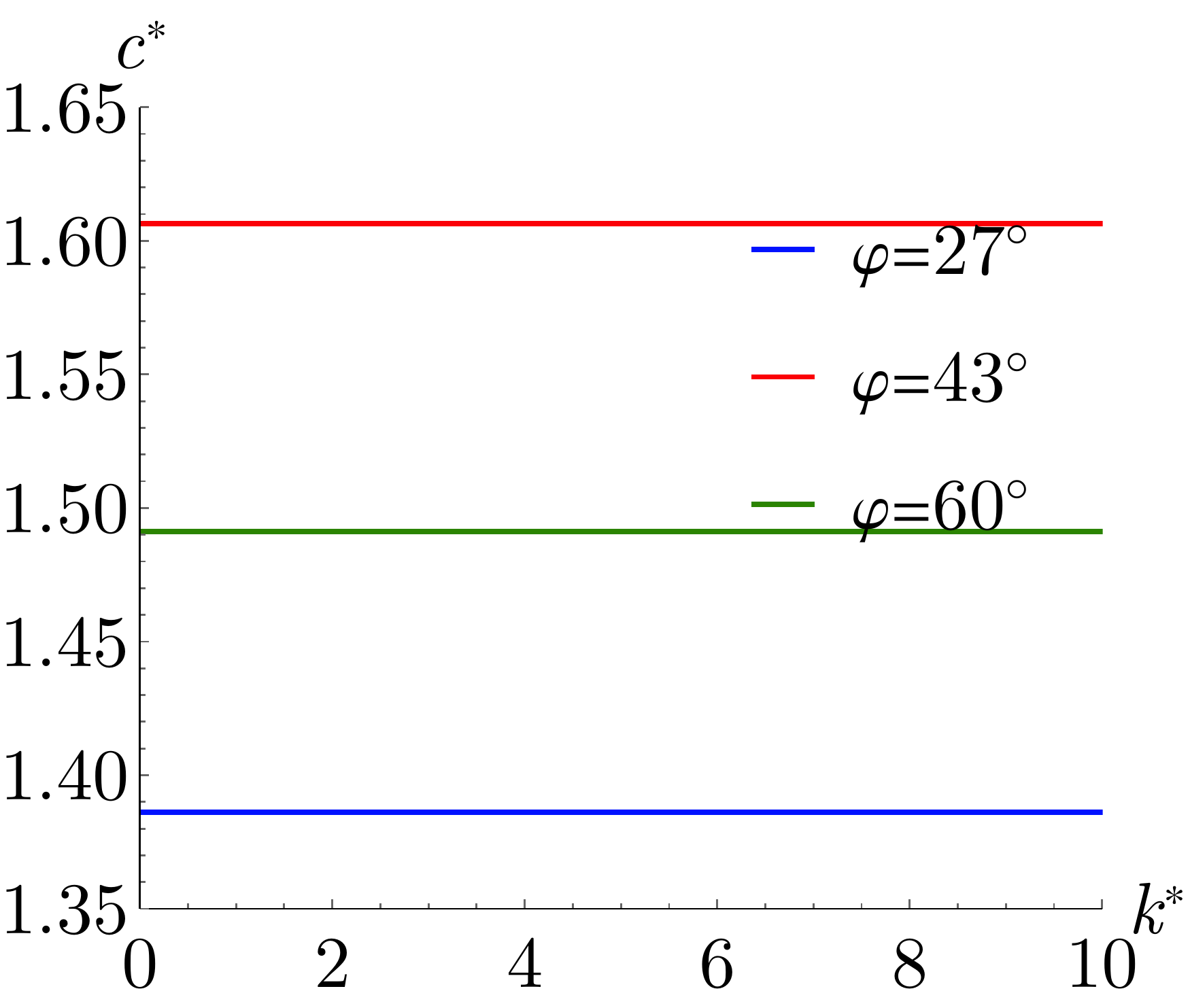}
		\end{minipage}
	}\qquad\quad\hspace{0.2em}
	\subfigure[]{
		\begin{minipage}{4.3cm}
			\centering
			\includegraphics[scale=0.29]{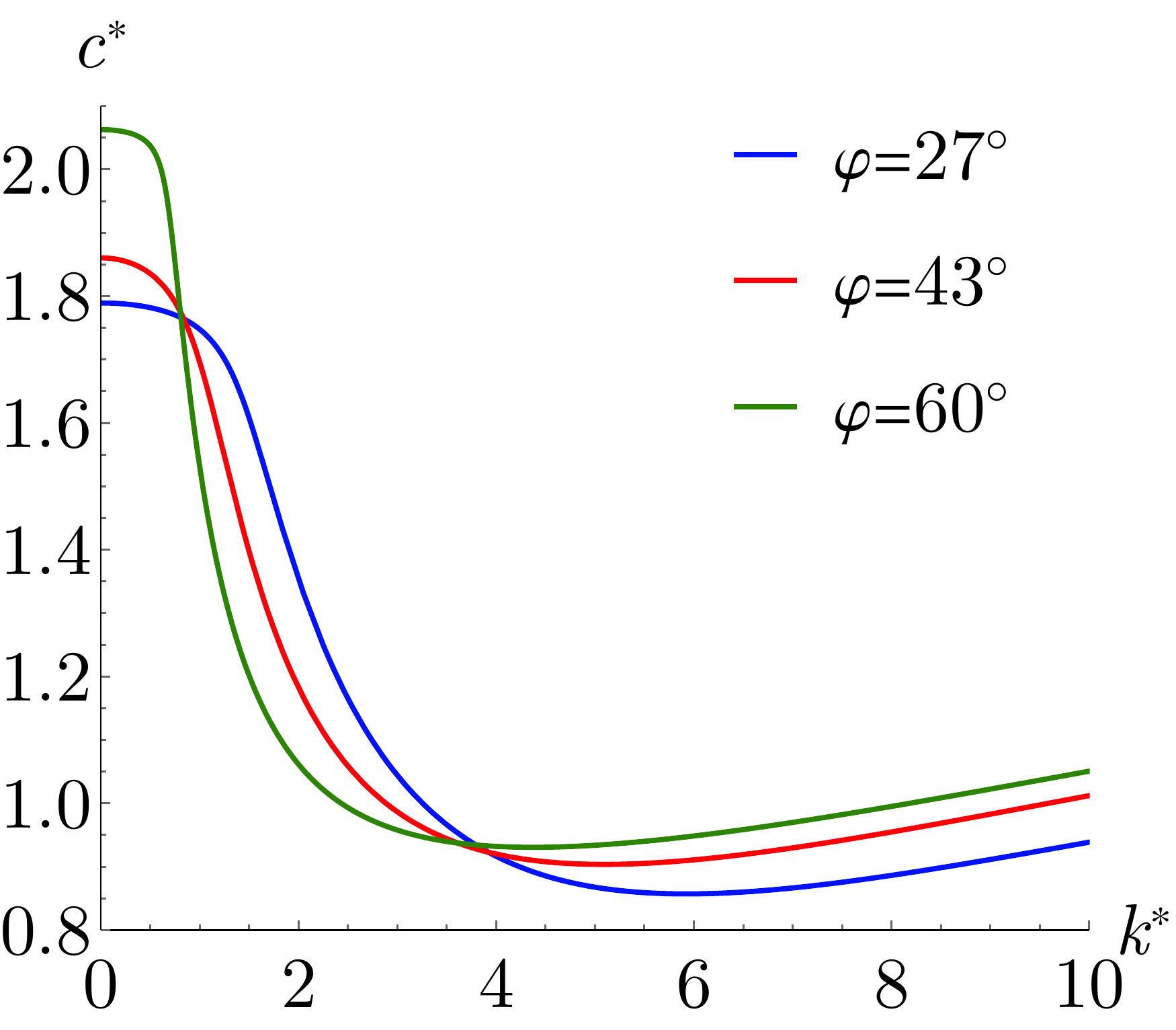}
		\end{minipage}
	}\qquad\quad\subfigure[]{
		\begin{minipage}{4.3cm}
			\centering
			\includegraphics[scale=0.29]{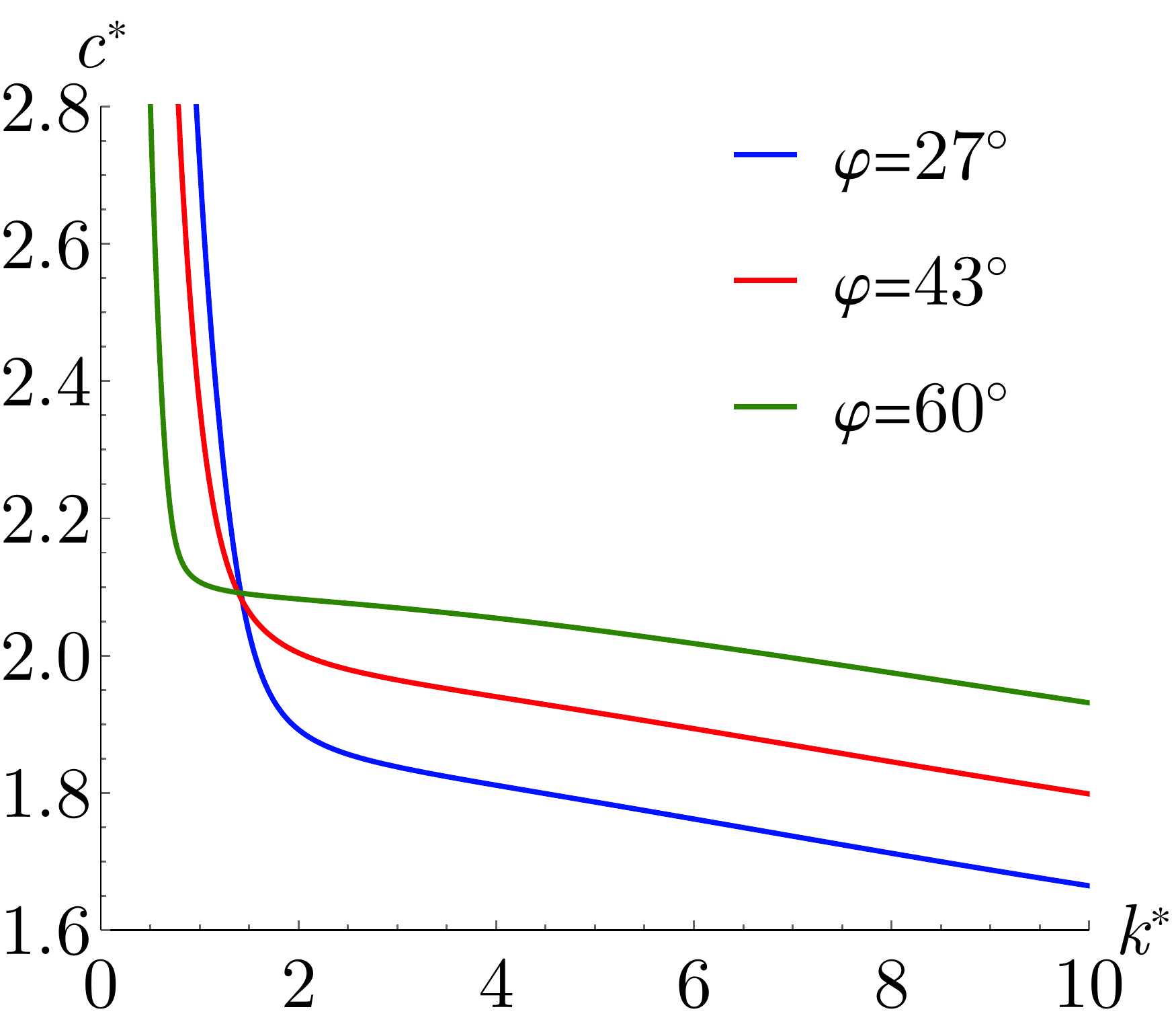}
		\end{minipage}
	}
	\caption{(Color online) Phase velocity spectra of axisymmetric waves at different fiber angles: (a) Circumferential wave; (b) Axial-radial wave; (c) Radial-axial wave.}
	\label{fig:d1}
\end{figure}

Finally, we examine the effect of fiber angle on the dispersion relations of axisymmetric waves.
For three different values of the fiber angle $\varphi=27^\circ$, $43^\circ$ and $60^\circ$ with $P^*=0.8$ and $\lambda_z=1$, the phase velocity spectra of  axisymmetric waves are displayed in Figure \ref{fig:d1}.
It is seen from Figure \ref{fig:d1} that the phase velocities of all the waves exhibit non-monotonic dependence on the fiber angle.

\subsection{Dispersion relations of non-axisymmetric waves}
We now investigate non-axisymmetric waves propagating in the pressurized artery for which $n\neq 0$. In the non-axisymmetric case, none of the $m_{ij}$, $1\leq i,j\leq 3$ in  \eqref{eq:Mm} vanishes. Consequently, the dispersion equation $\det(M)=0$ represents motions where displacements in the circumferential, axial and radial directions are all coupled. We shall focus on the non-axisymmetric waves with the lowest circumferential wavenumber $n=1$, which are also known as the fundamental flexural waves. As in the axisymmetric case, we can divide the non-axisymmetric waves with $n=1$ into three types which are called the {\it circumferential-radial-axial wave},  {\it axial-radial-circumferential wave} and {\it radial-circumferential-axial wave}, respectively, depending on their displacement components in the long wavelength limit $k\to 0$. Specifically, when $k=0$, among the three displacement components of the circumferential-radial-axial wave, the (magnitude of) circumferential component is the largest, the radial component is the next largest and the axial component is the smallest; similar relations among the displacement components apply to  the other two waves. In \cite{Gaza}, these non-axisymmetric waves are also called the {\it flexural waves}.

\subsubsection{Validation of the present theory with non-axisymmetric waves}

\begin{figure}[h]
	\centering
	\hspace{-2em}	\subfigure[]{
		\begin{minipage}{4.3cm}
			\centering
			\includegraphics[scale=0.29]{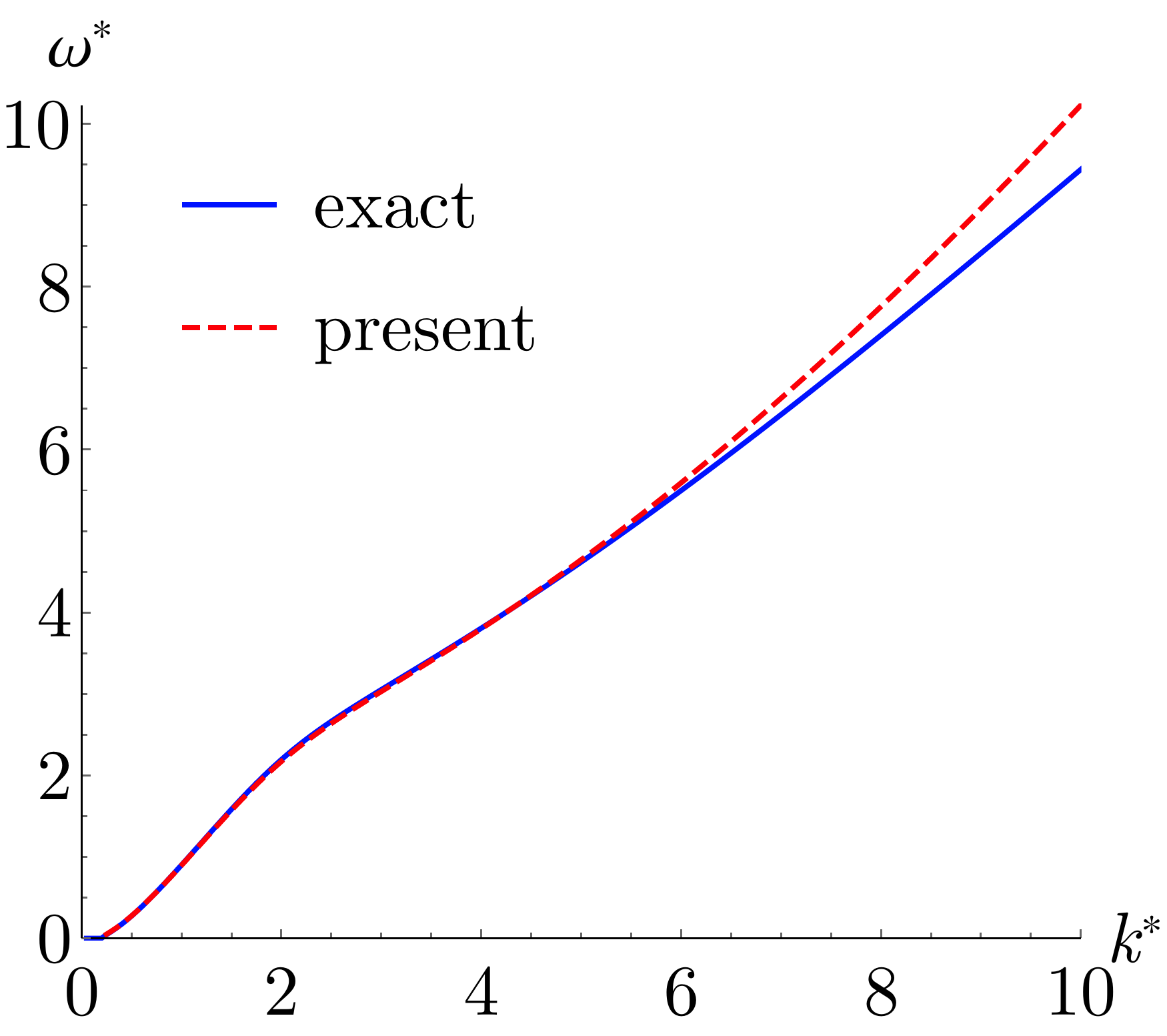}
		\end{minipage}
	}\qquad\quad
	\subfigure[]{
		\begin{minipage}{4.3cm}
			\centering
			\includegraphics[scale=0.29]{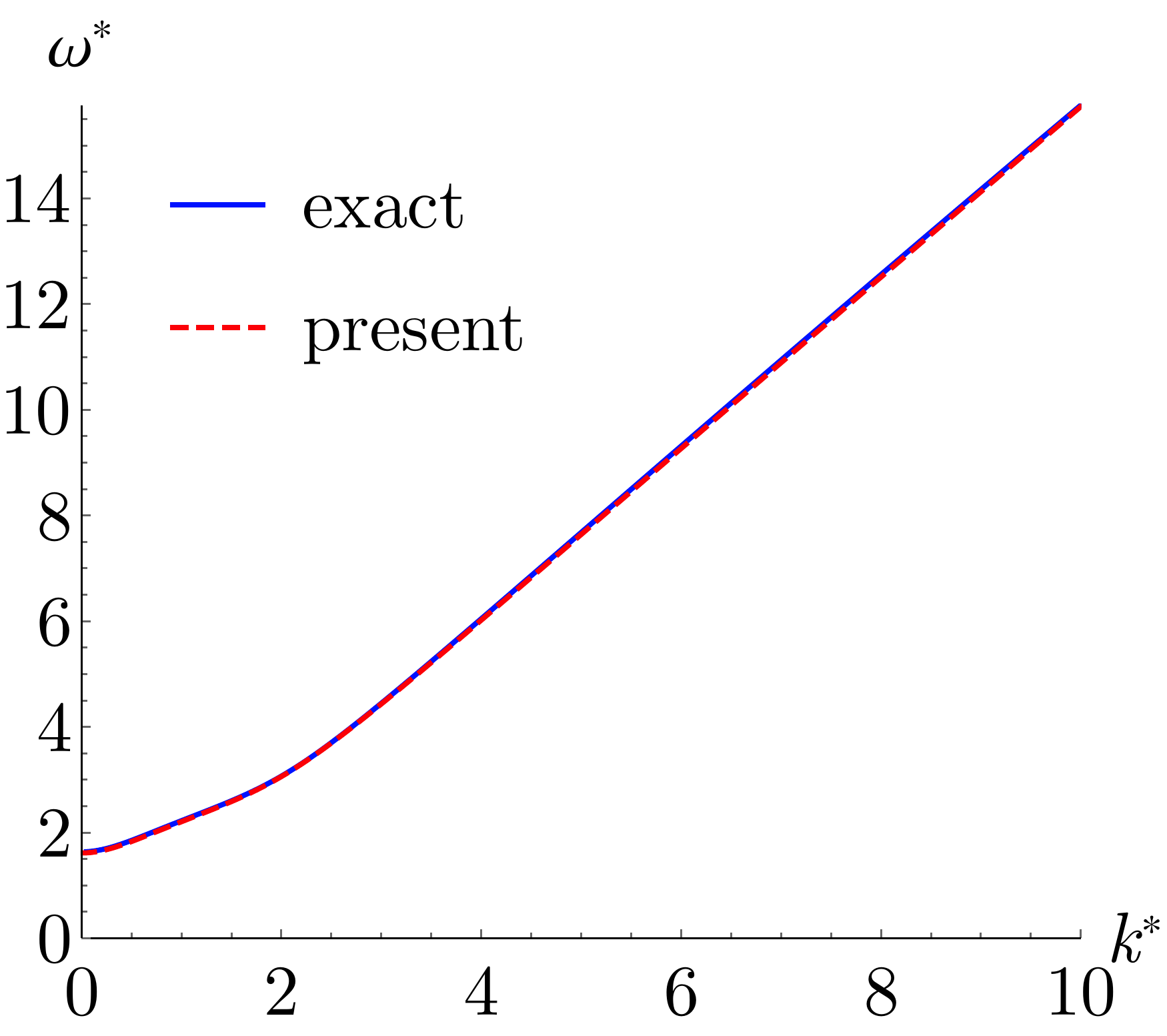}
		\end{minipage}
	}\qquad\quad\subfigure[]{
		\begin{minipage}{4.3cm}
			\centering
			\includegraphics[scale=0.29]{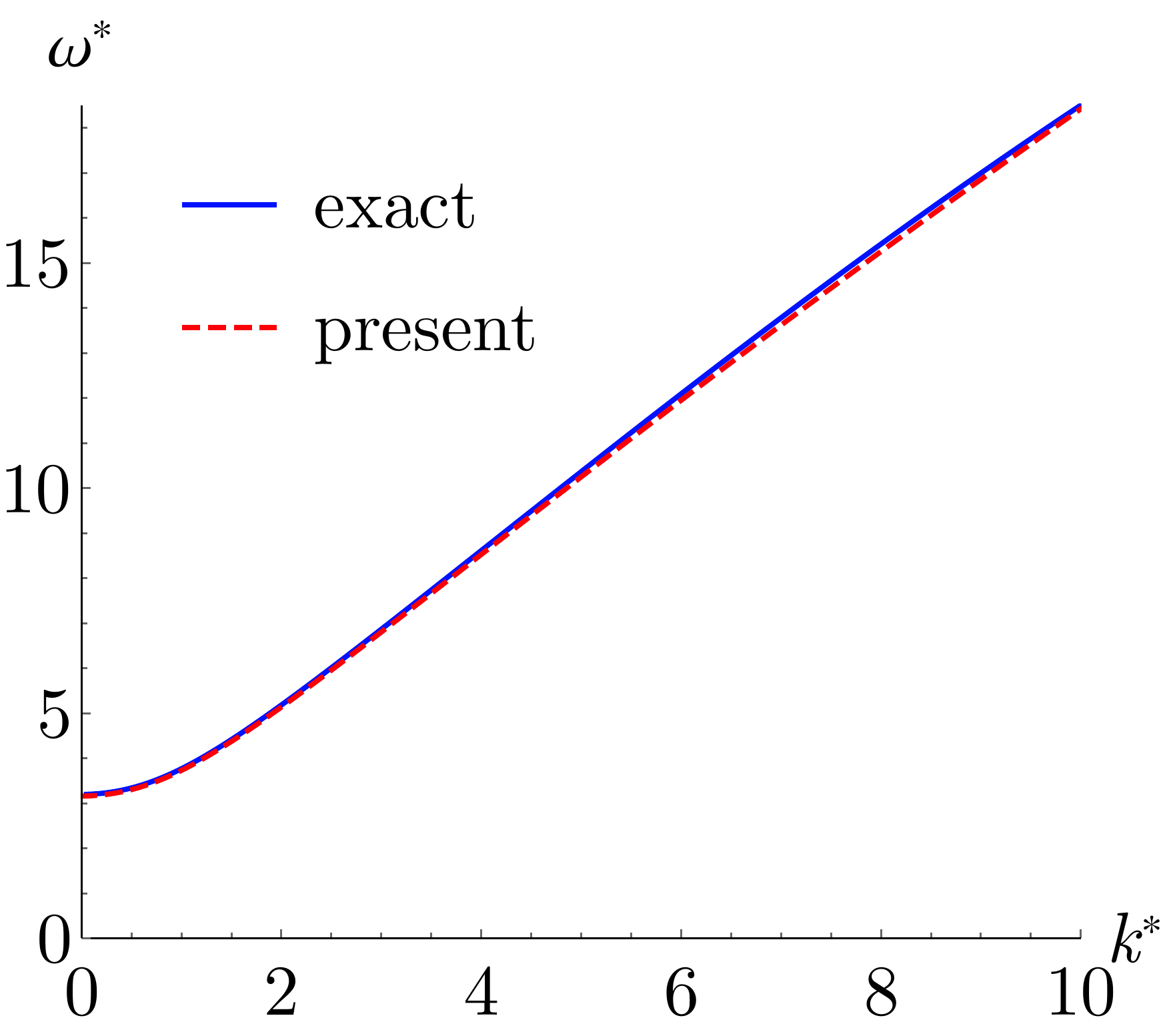}
		\end{minipage}
	}
	\caption{(Color online) Comparison of the frequency spectra of non-axisymmetric waves with $n=1$ obtained by the exact theory and present theory: (a) Circumferential-radial-axial wave;  (b) Axial-radial-circumferential wave;  (c) Radial-circumferential-axial wave. }
	\label{fig:na1}
\end{figure}

\begin{figure}[h]
	\centering
	\hspace{-2.8em}	\subfigure[]{
		\begin{minipage}{4.3cm}
			\centering
			\includegraphics[scale=0.29]{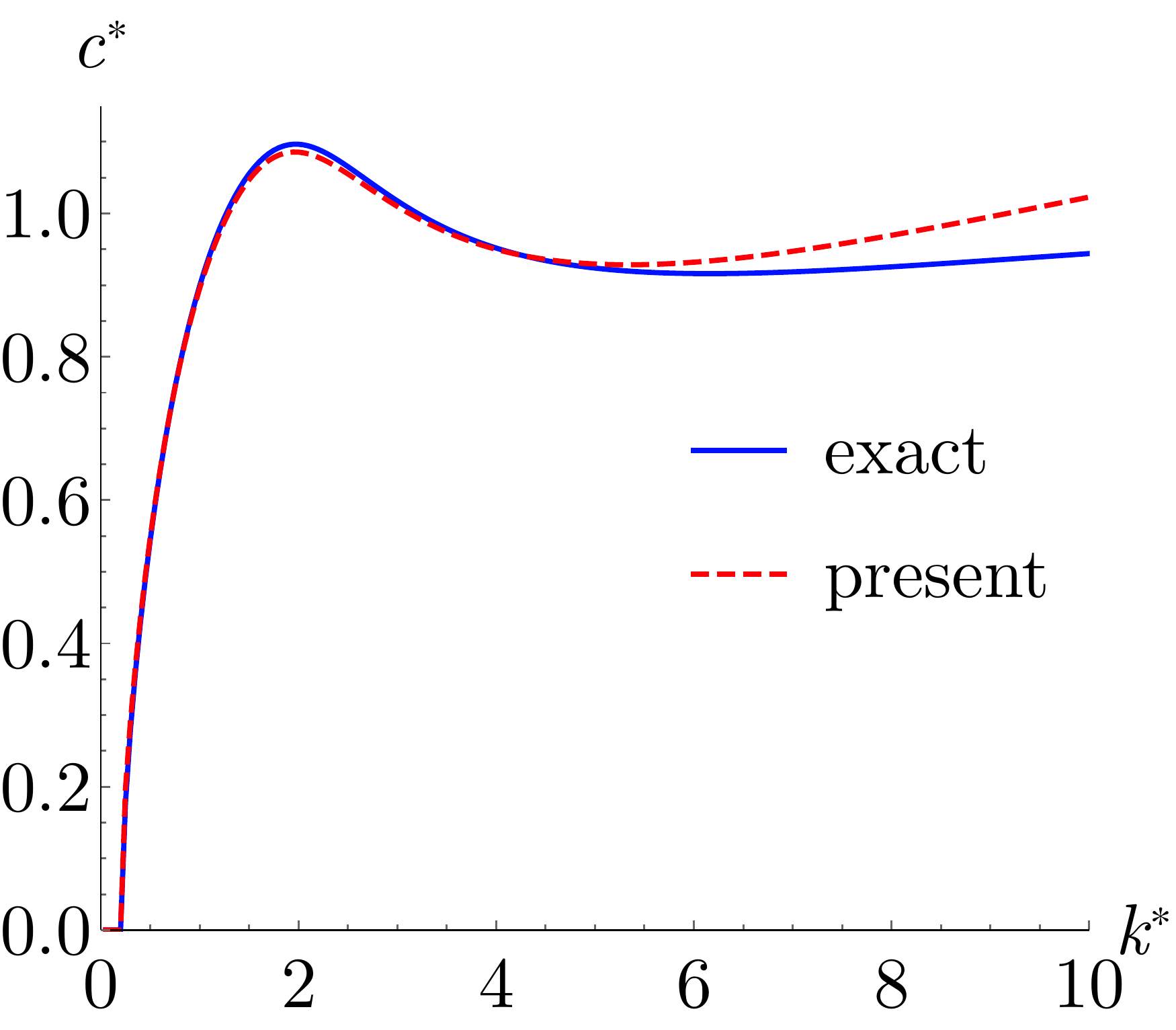}
		\end{minipage}
	}\qquad\quad\hspace{-0.8em}
	\subfigure[]{
		\begin{minipage}{4.3cm}
			\centering
			\includegraphics[scale=0.29]{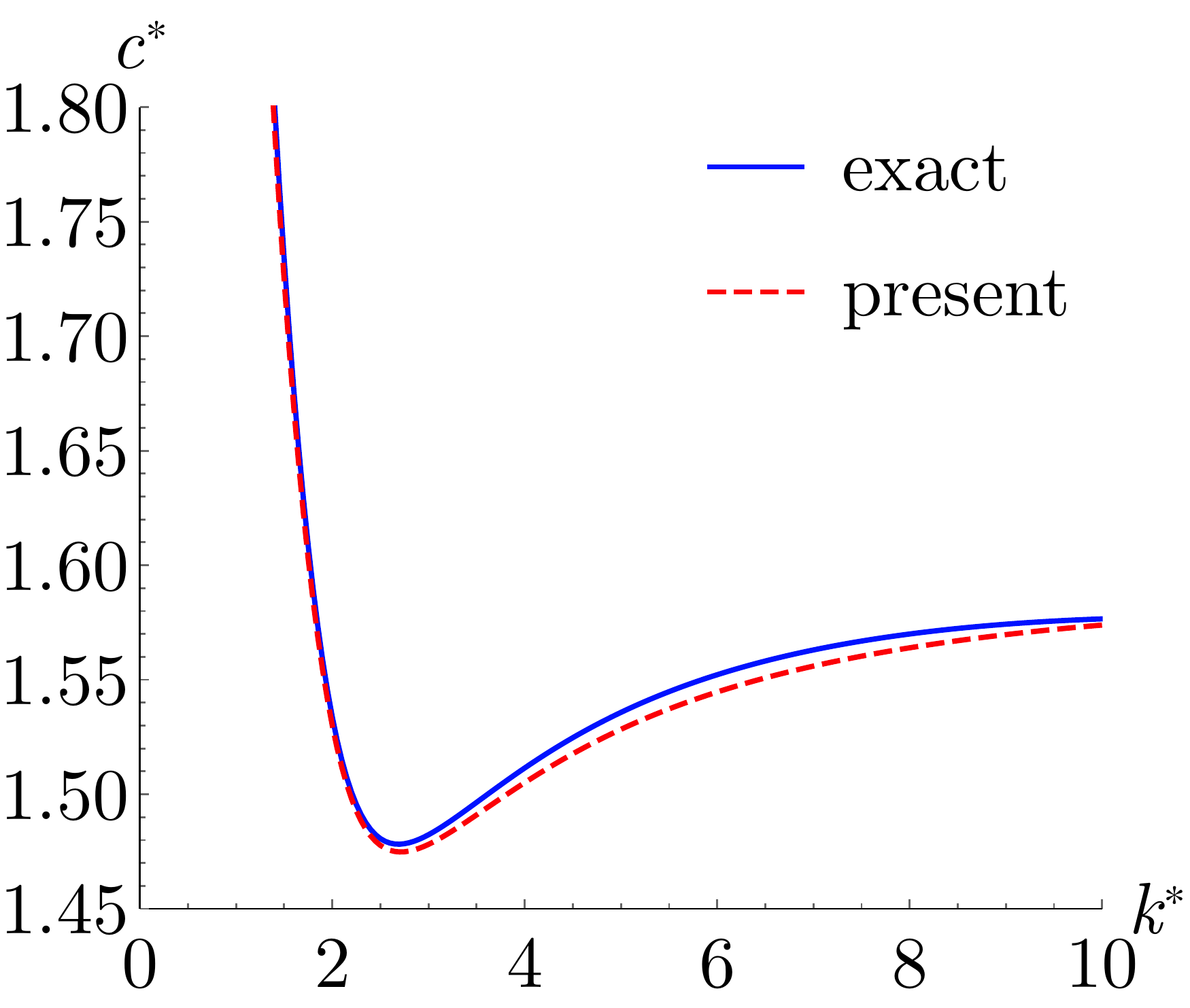}
		\end{minipage}
	}\qquad\quad\subfigure[]{
		\begin{minipage}{4.3cm}
			\centering
			\includegraphics[scale=0.29]{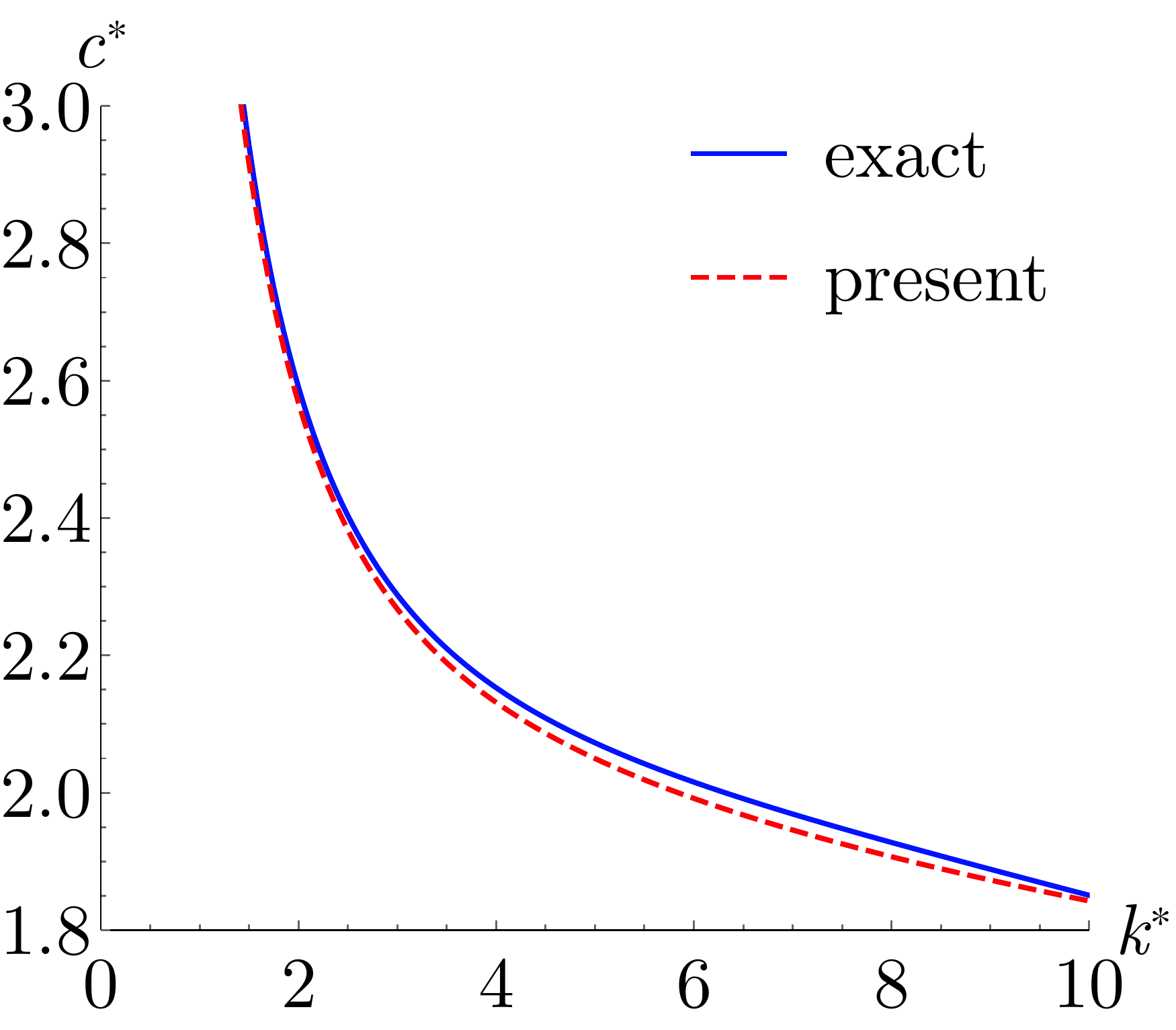}
		\end{minipage}
	}
     \caption{(Color online) Comparison of the phase velocity spectra of non-axisymmetric waves with $n=1$ obtained by the exact theory and present theory: (a) Circumferential-radial-axial wave;  (b) Axial-radial-circumferential wave;  (c) Radial-circumferential-axial wave. }
     \label{fig:nb1}
\end{figure}

As before, we first validate the present theory with the dispersion relations of non-axisymmetric waves. For fixed $P^*=0.8$ and $\lambda_z=1$, the frequency spectra and phase velocity spectra of non-axisymmetric waves with $n=1$ based on the exact theory and present theory are shown in Figure \ref{fig:na1} and Figure \ref{fig:nb1}, respectively. It is seen from the figures the dispersion curves obtained by the present theory are almost indistinguishable with the exact dispersion curves when  the wavenumber is small (say, $k^*\leq 3$ which is equivalent to $kh\leq 0.3$), validating that the present theory is capable of delivering asymptotically correct results. When the wavenumber becomes large, the agreement between the present theory and exact theory becomes poor, but the present theory is still able to capture the correct shapes of the phase velocity  curves, as shown in Figure \ref{fig:nb1}.

\subsubsection{Effect of the pressure}
\begin{figure}[h]
	\centering
	\hspace{-2em}	\subfigure[]{
		\begin{minipage}{4.3cm}
			\centering
			\includegraphics[scale=0.29]{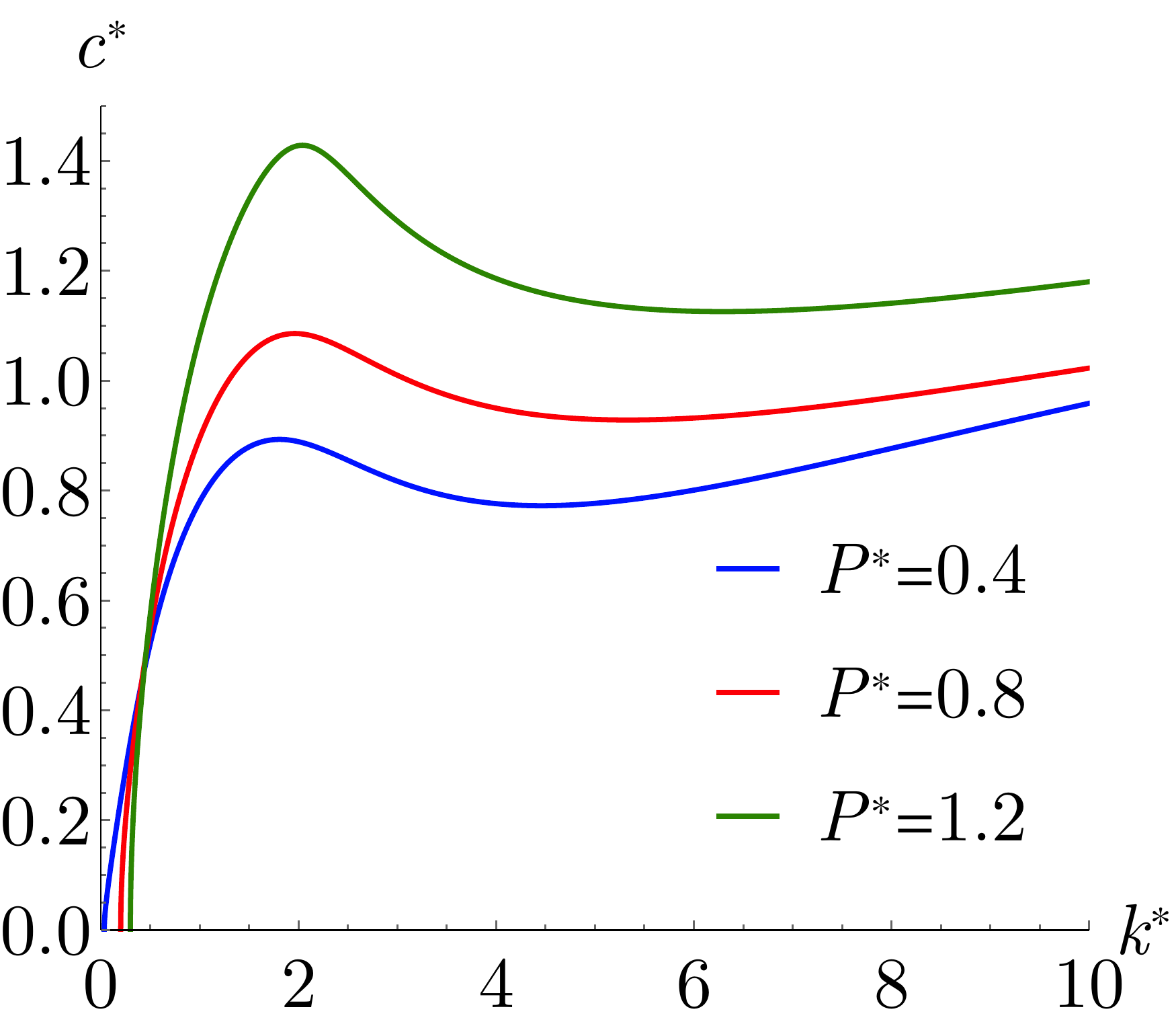}
		\end{minipage}
	}\qquad\quad
	\subfigure[]{
		\begin{minipage}{4.3cm}
			\centering
			\includegraphics[scale=0.29]{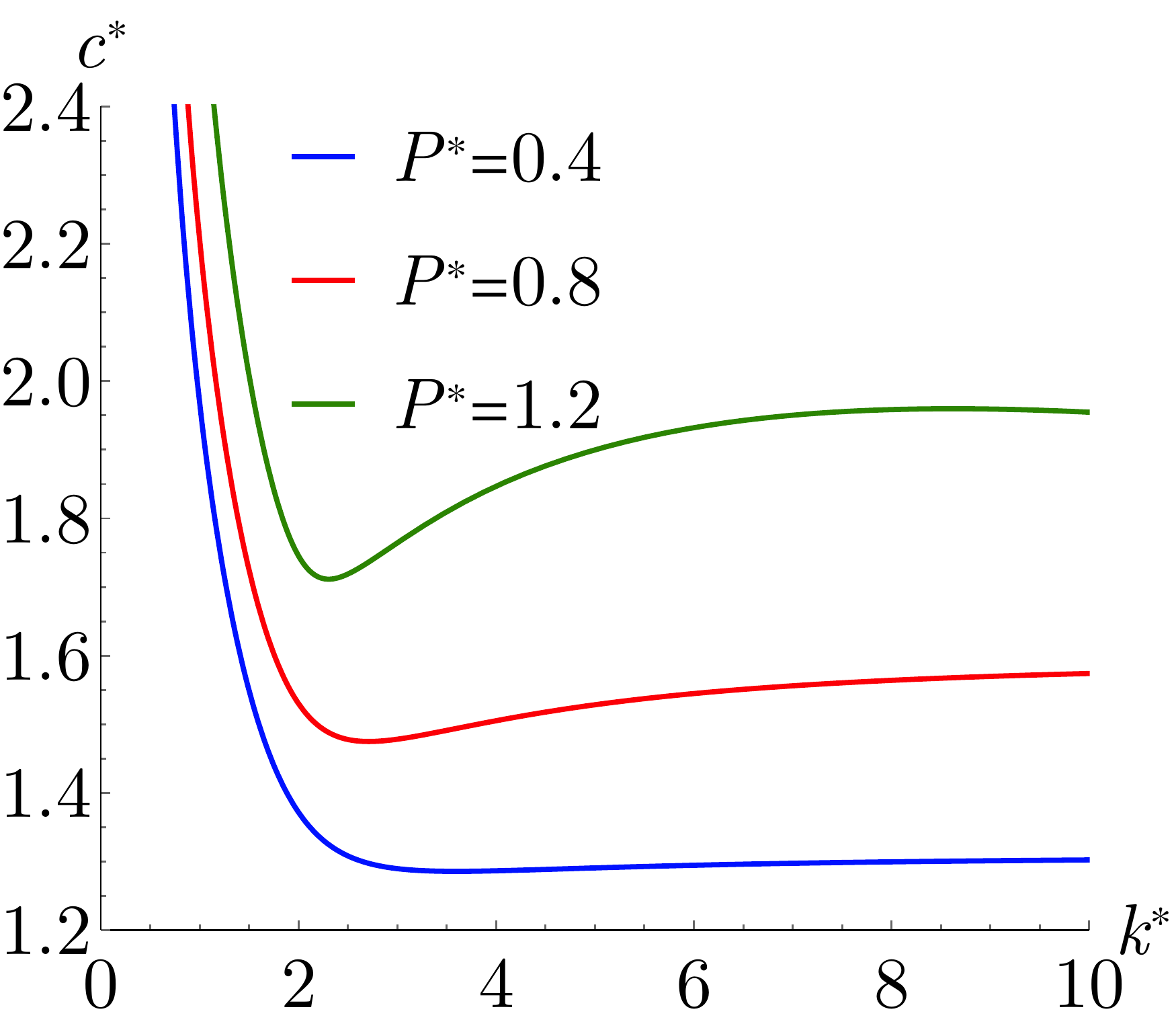}
		\end{minipage}
	}\qquad\quad\subfigure[]{
		\begin{minipage}{4.3cm}
			\centering
			\includegraphics[scale=0.29]{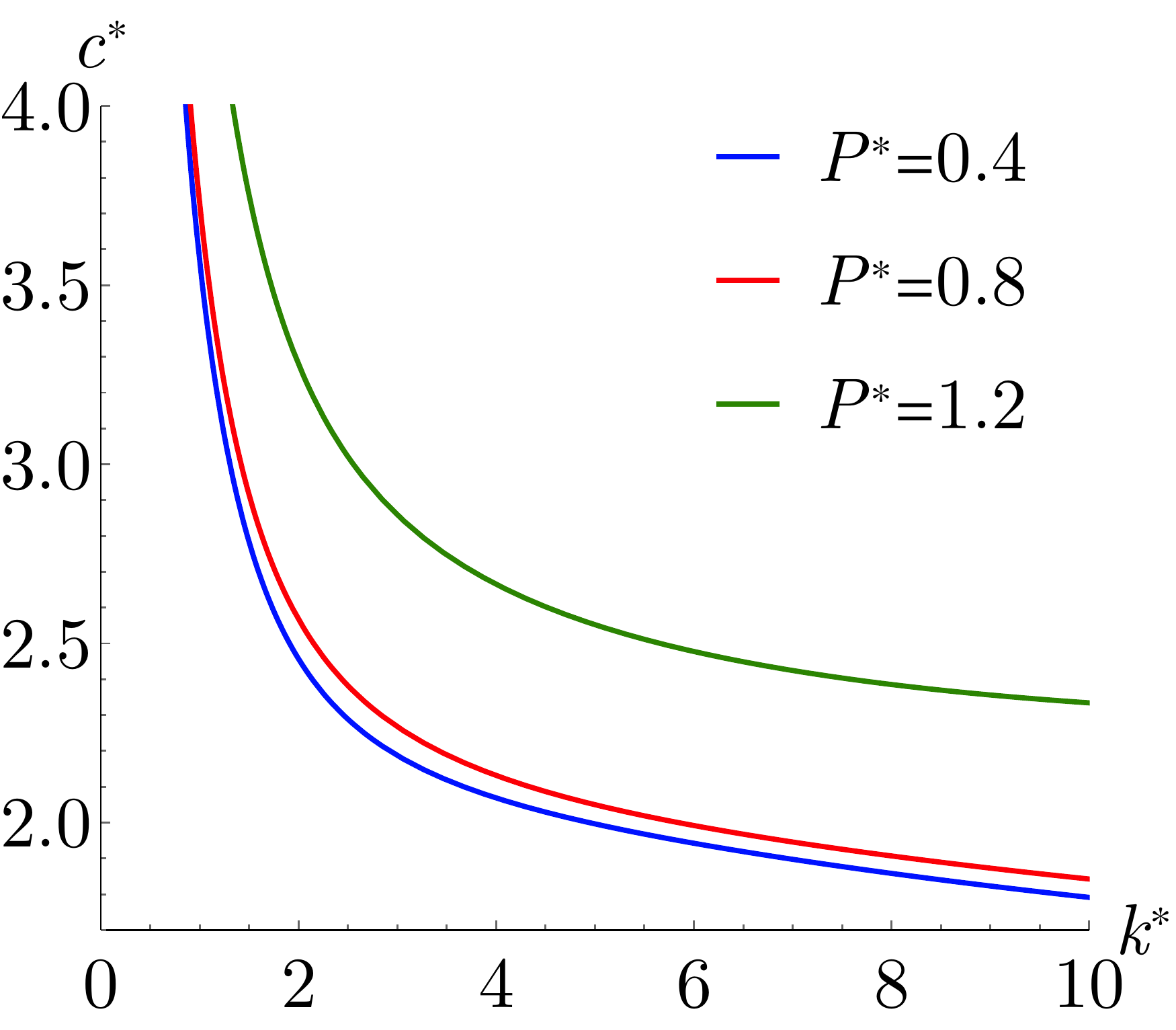}
		\end{minipage}
	}
	\caption{(Color online) Phase velocity spectra of non-axisymmetric waves with $n=1$ at different pressures: (a) Circumferential-radial-axial wave;  (b) Axial-radial-circumferential wave;  (c) Radial-circumferential-axial wave. }
	\label{fig:nc1}
\end{figure}

 Now we examine the influence of the pressure on the dispersion relations of non-axisymmetric waves with $n=1$.  For  three different values of the pressure $P^*=0.4$, $0.8$ and $1.2$ with $\lambda_z=1$,  the phase velocity spectra are shown in Figure \ref{fig:nc1}. It is seen from Figure \ref{fig:nc1} that all the waves are dispersive. Also,
 the phase velocities of all the waves increase with the pressure in the entire wavenumber range considered except that of the circumferential-radial-axial wave in the small wavenumber region. In addition, it is observed that the circumferential-radial-axial wave does not exist in the small wavenumber region when the pressure becomes relatively large, say $P^*\geq 0.8$. Besides, it is noticed that the phase velocity of the radial-circumferential-axial wave decreases monotonically  as the wavenumber increases in all the cases, while those of circumferential-radial-axial and axial-radial-circumferential waves in all the cases vary non-monotonically with the wavenumber.

\subsubsection{Effects of the axial pre-stretch and fiber angle}

\begin{figure}[h]
	\centering
	\hspace{-2em}	\subfigure[]{
		\begin{minipage}{4.3cm}
			\centering
			\includegraphics[scale=0.29]{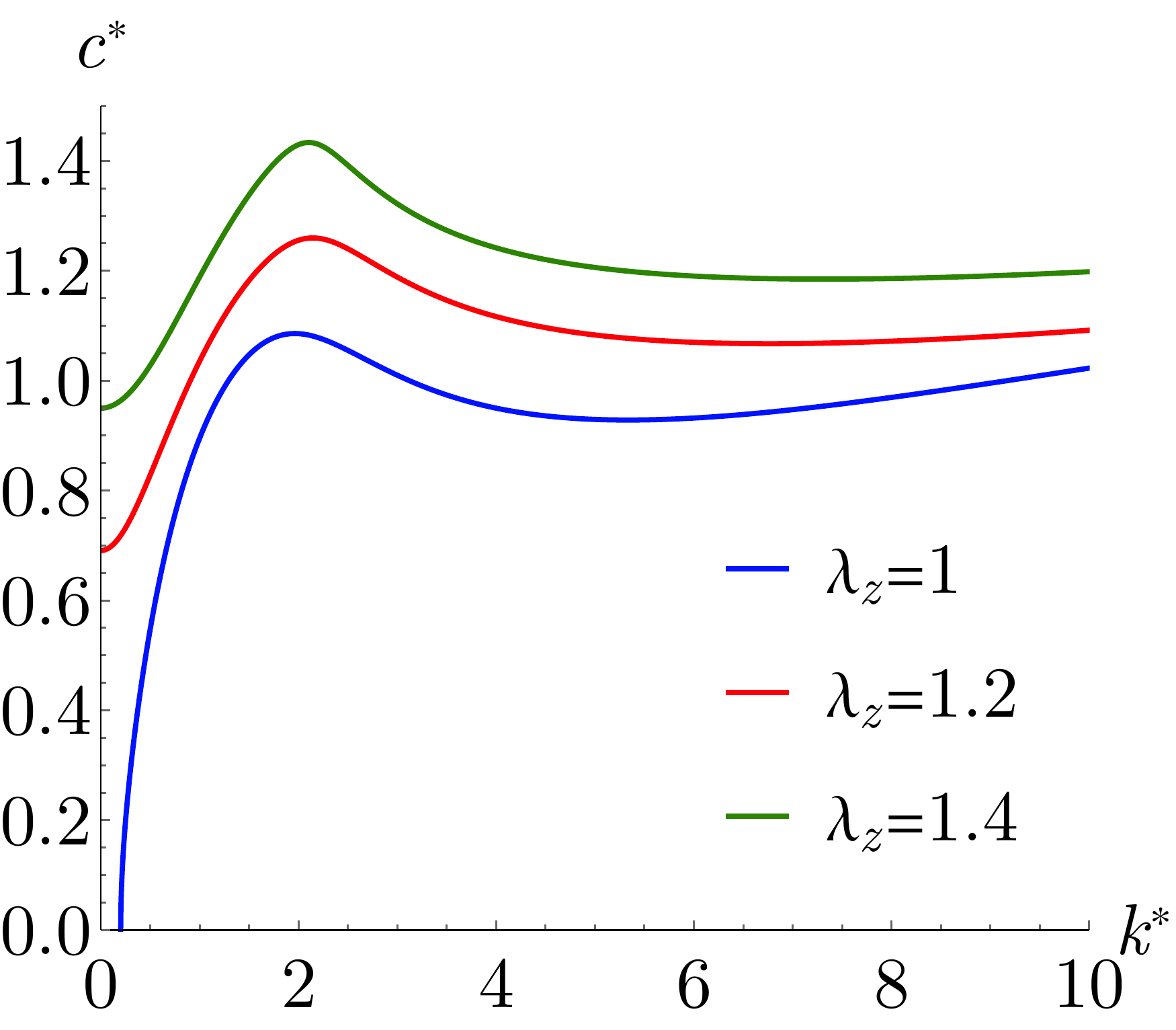}
		\end{minipage}
	}\qquad\quad
	\subfigure[]{
		\begin{minipage}{4.3cm}
			\centering
			\includegraphics[scale=0.29]{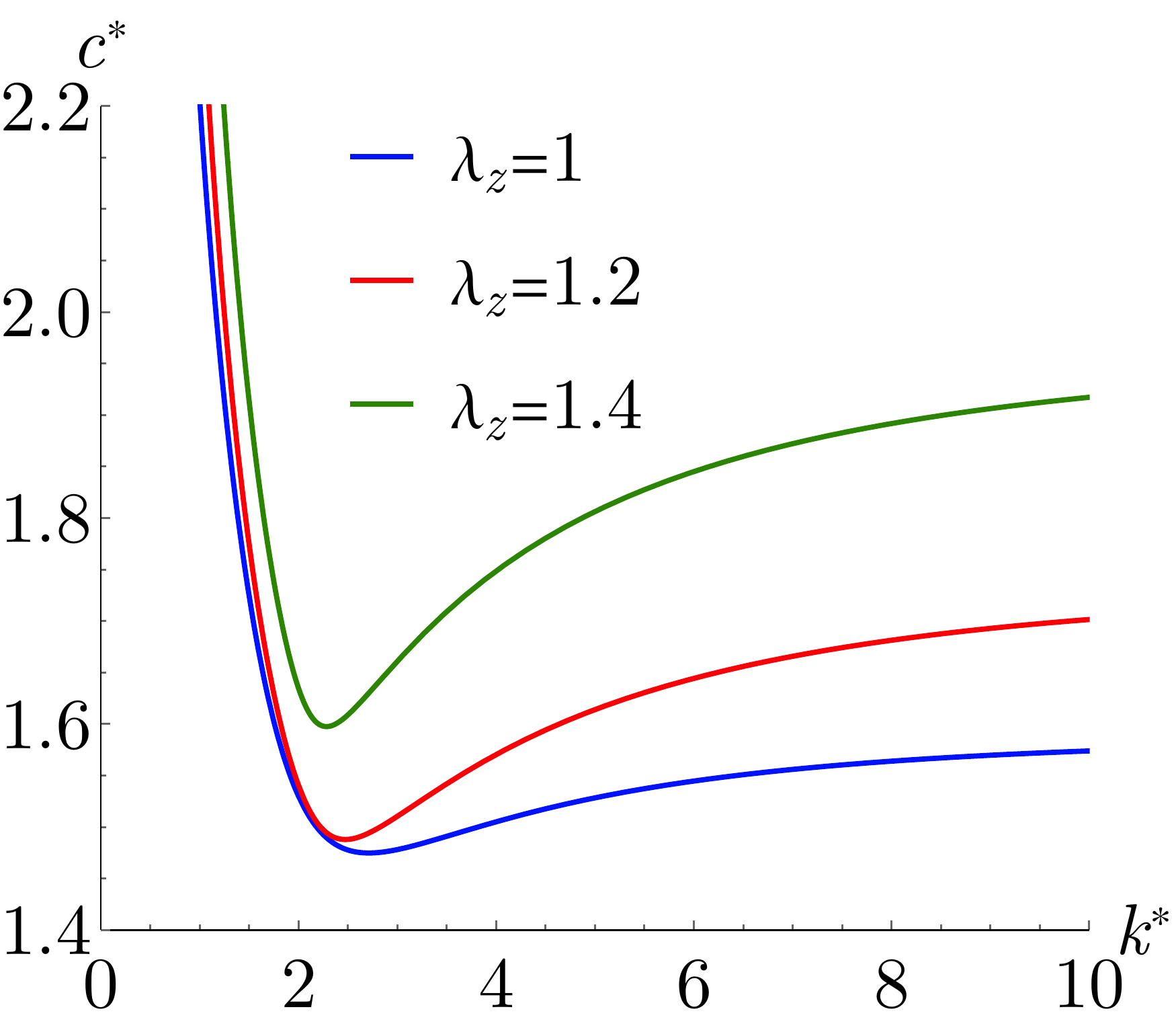}
		\end{minipage}
	}\qquad\quad\subfigure[]{
		\begin{minipage}{4.3cm}
			\centering
			\includegraphics[scale=0.29]{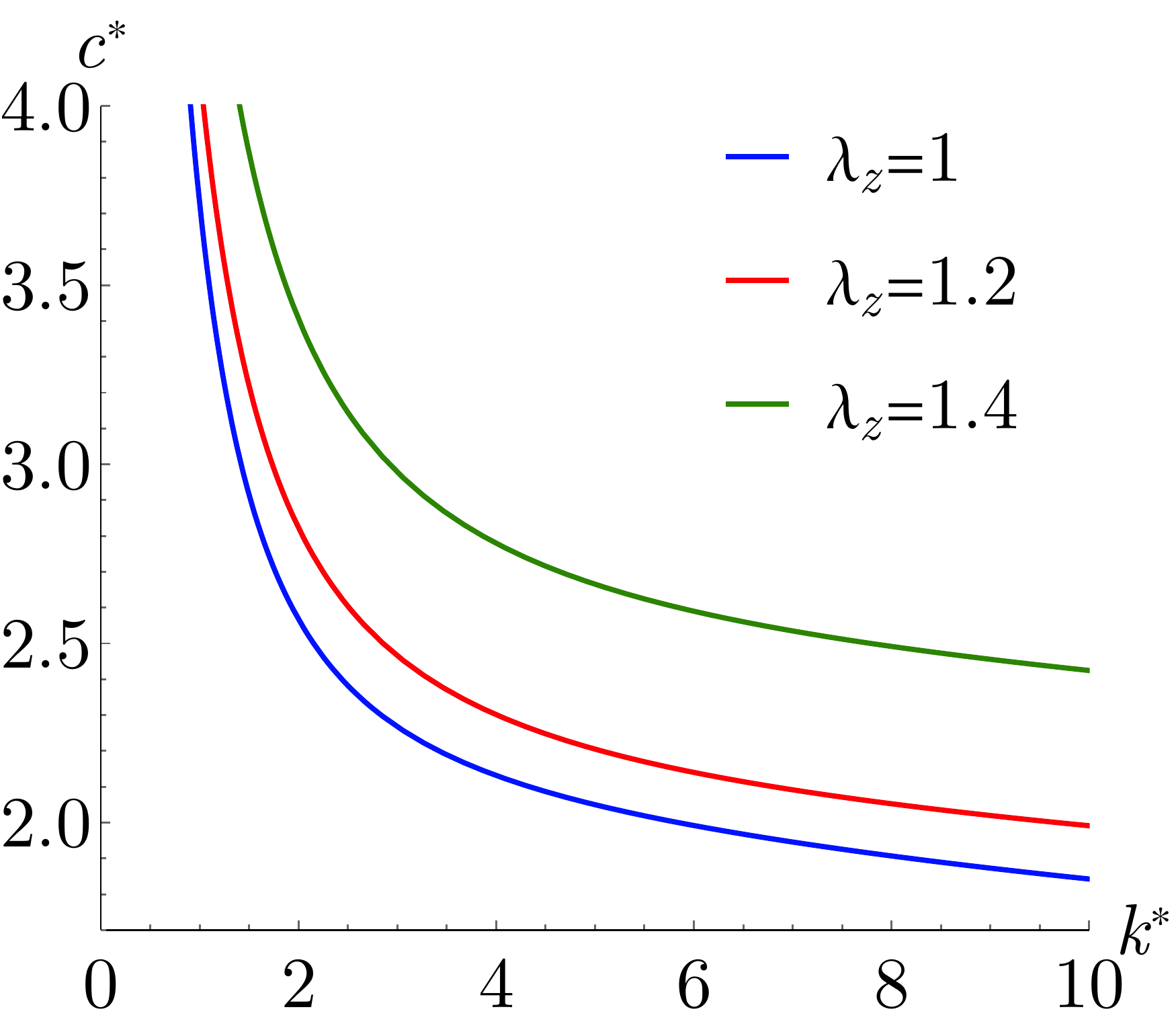}
		\end{minipage}
	}
	\caption{(Color online) Phase velocity spectra of non-axisymmetric waves with $n=1$ at different axial pre-stretches:  (a) Circumferential-radial-axial wave;  (b) Axial-radial-circumferential wave;  (c) Radial-circumferential-axial wave. }
	\label{fig:nd1}
\end{figure}

Then we turn to study how the axial pre-stretch affects the dispersion relations of non-axisymmetric waves with $n=1$. For three different values of the axial pre-stretch $\lambda_z=1, 1.2, 1.4$ with $P^*=0.8$, the phase velocity spectra is shown in Figure \ref{fig:nd1}. It is seen from Figure \ref{fig:nd1} that the phase velocities of all the waves increase with the axial pre-stretch in the entire wavenumber range shown.

\begin{figure}[h]
	\centering
	\hspace{-2em}	\subfigure[]{
		\begin{minipage}{4.3cm}
			\centering
			\includegraphics[scale=0.29]{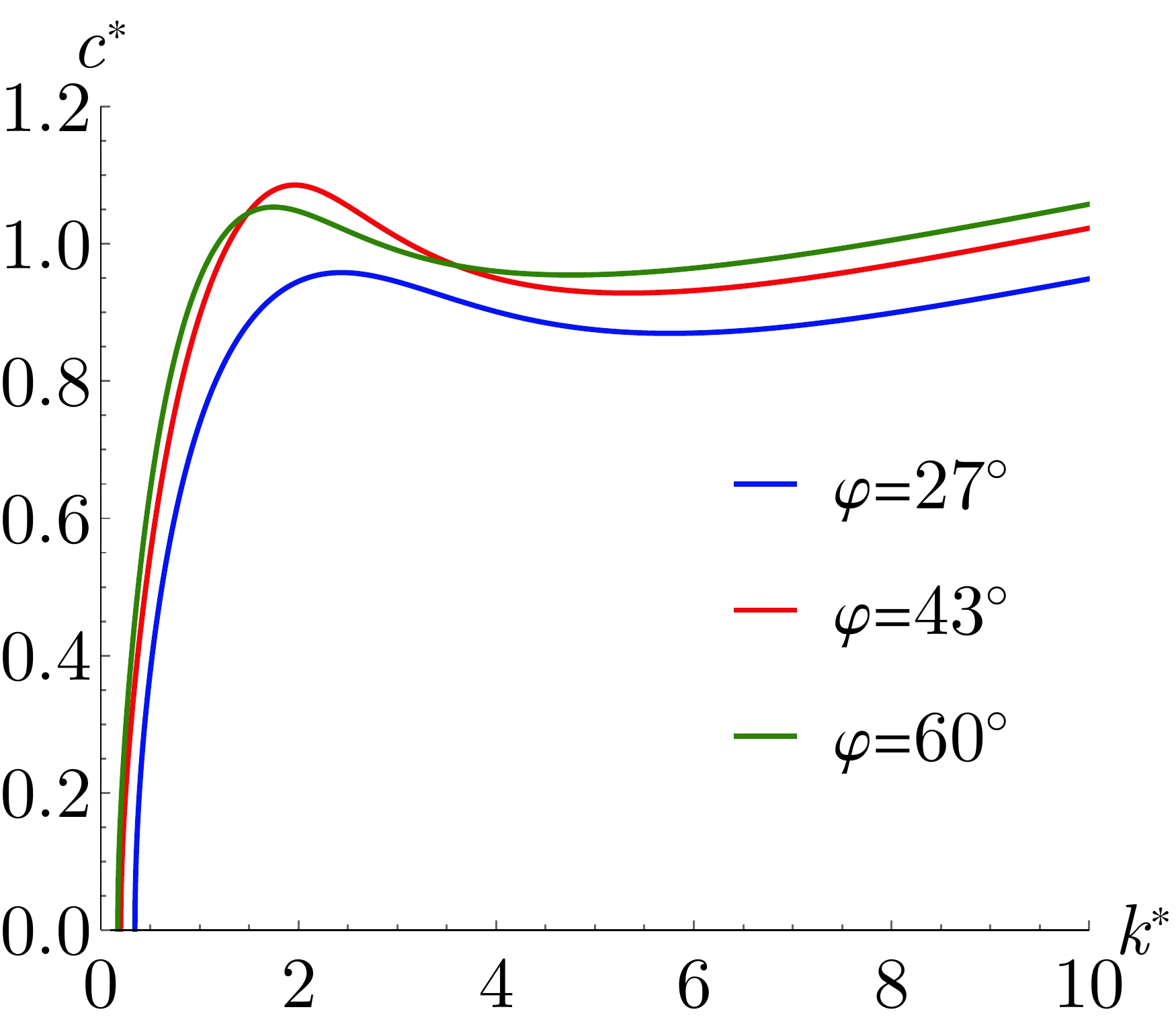}
		\end{minipage}
	}\qquad\quad
	\subfigure[]{
		\begin{minipage}{4.3cm}
			\centering
			\includegraphics[scale=0.29]{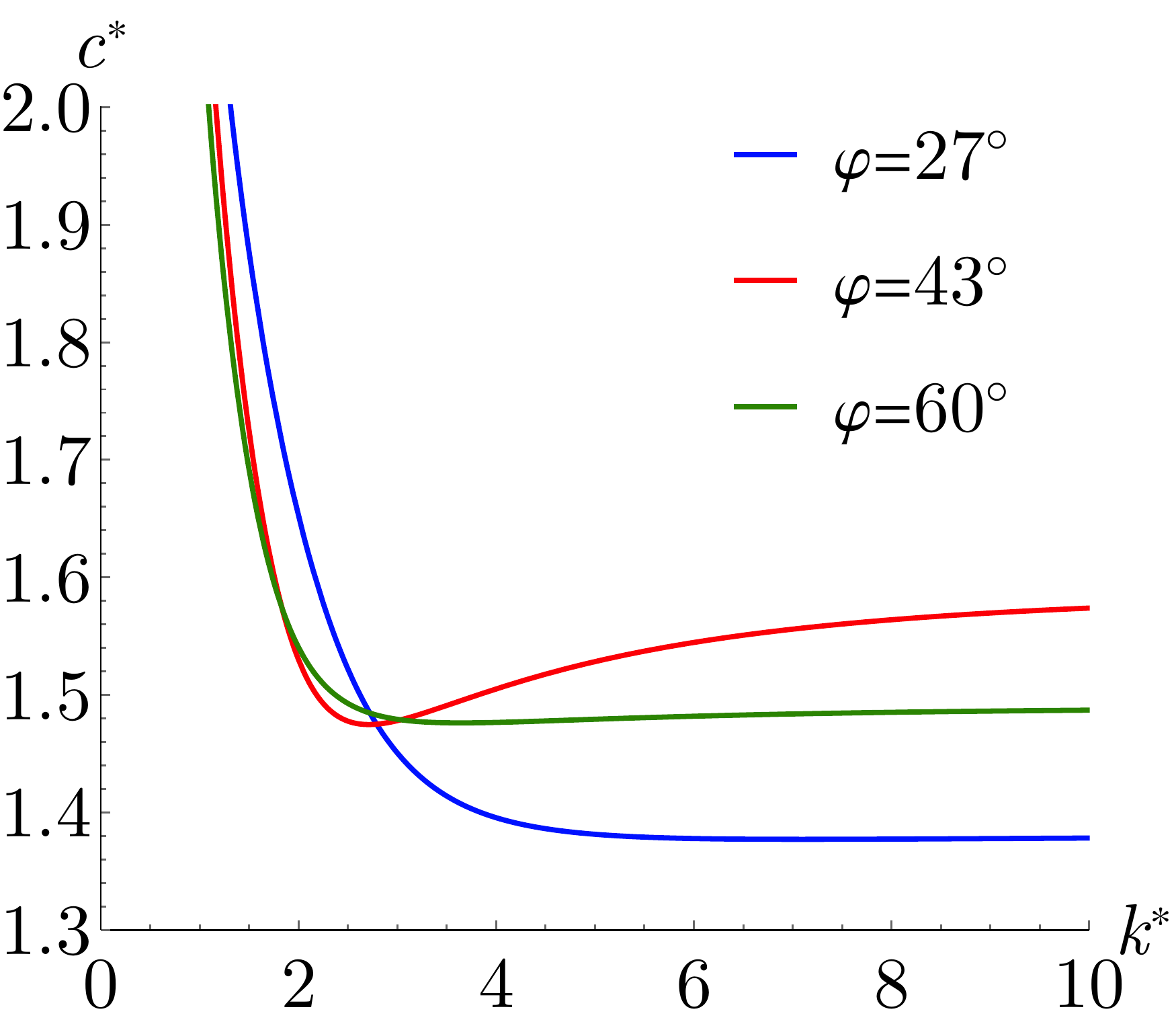}
		\end{minipage}
	}\qquad\quad\subfigure[]{
		\begin{minipage}{4.3cm}
			\centering
			\includegraphics[scale=0.29]{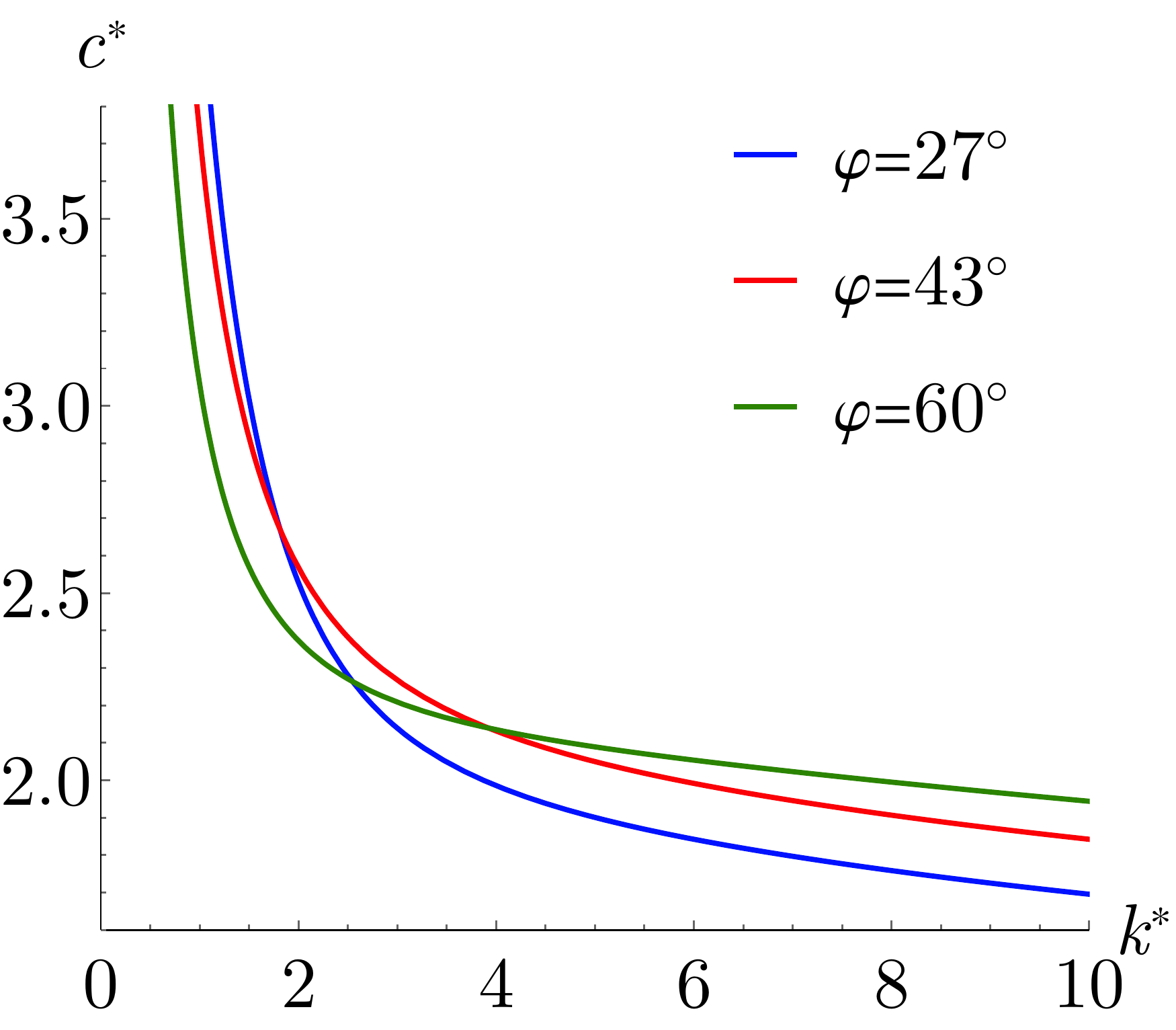}
		\end{minipage}
	}
	\caption{(Color online) Phase velocity spectra of non-axisymmetric waves with $n=1$ at different fiber angles:  (a) Circumferential-radial-axial wave;  (b) Axial-radial-circumferential wave;  (c) Radial-circumferential-axial wave. }
	\label{fig:ne1}
\end{figure}

Finally, we examine how the fiber angle affects the dispersion relations of non-axisymmetric waves with $n=1$. For three different values of the fiber angle $\varphi=27^\circ$, $43^\circ$ and $60^\circ$ with $P^*=0.8$ and $\lambda_z=1$, the  phase velocity spectra  are displayed Figure \ref{fig:ne1}. It is found that similar to the axisymmetric case, the phase velocities of all the waves demonstrate exhibit complex non-monotonic variations with the fiber angle.

\section{Conclusion}
In this paper, we investigated both axisymmetric and non-axisymmetric waves in a  fiber-reinforced hyperelastic tube under internal pressure. Once the material model is specified, the exact dispersion relation in each case can only be obtained by solving a two-point boundary value problem numerically. On the other hand, although the corresponding dispersion relations based on the membrane assumption can be derived in closed form, they cannot describe the effect of bending stiffness. The reduced model developed in the present paper provides a compromise: it takes into account the effect of bending stiffness and can be used to derive the dispersion relations analytically. Our main aim has been to assess its accuracy by comparing its predictions with those given by the exact theory, and to give additional results on the effects of axial pre-stretch, internal pressure and fiber orientation angle on the dispersion curves.

It is shown that the present theory is sufficiently accurate for studying wave propagation and its relative errors for the dispersion curves of axisymmetric waves in a pressurized artery are less than $1.9\%$ for wavenumbers with $kh\leq 0.6$ and less than $8.2\%$ in the whole wavenumber range considered where $kh\leq 1$. By comparing the dispersion curves by the present theory and membrane theory, it is found that the present theory is  more accurate than the membrane theory and the bending effect plays an important role in studying axisymmetric wave modes for relatively large wavenumbers, especially in studying the radial-axial wave mode. It is hoped that the proposed methodology can be used to deduce material properties from measurements of wave characteristics.

\section*{Acknowledgments}
The work described in this paper is fully supported by a GRF grant (Project no. CityU 11303718) from the Research Grants Council of the Government of HKSAR, China.

\appendix

\section{The recurrence relations and refined shell equations for a neo-Hookean cylindrical tube}\label{app:neo}

As an illustrative example, we may specify the recurrence relations to the shell that is a cylindrical tube in the reference configuration and made of the neo-Hookean material. Thus the strain energy function is of the form $W(\bm{F})=\frac{\mu}{2}(\tr(\bm{F}^T\bm{F})-3)$, where $\mu$ is the shear modulus. We assume that the tube undergoes an axisymmetric deformation described by $\bm{x}(\bm{X})=r(R, X)\bm{e}_R+z(R, X)\bm{e}_{X}$, where $R$ and $X$ represent respectively the radial and axial coordinates with the associated unit vectors  $\bm{e}_R$ and $\bm{e}_X$.  In this case the recurrence relations \eqref{eq:A00} and \eqref{eq:R00} become
\begin{align}
&\mu z_1+p_0\frac{r_0r'_0}{R_m}=m_X,\quad \mu r_1-p_0\frac{r_0z'_0}{R_m}=m_R,\quad z'_0r_1-r'_0z_1=\frac{R_m}{r_0},
\end{align}
and the recurrence relations \eqref{eq:x2} and \eqref{eq:p1} become
\begin{align}
&\mu z_2+p_1\frac{r_0r'_0}{R_m}+\mu(z''_0+\frac{z_1}{R_m})-p'_0\frac{r_0r_1}{R_m}=0,\\
& \mu r_2-p_1\frac{r_0z'_0}{R_m}+\mu(r''_0+\frac{r_1R_m-r_0}{R_m^2})+p'_0\frac{r_0z_1}{R_m}=0,\\
&z'_0r_2-r'_0z_2+z'_1r_1-z_1r'_1=\frac{r_0-R_mr_1}{r_0^2},
\end{align}
where $r_i$ and $z_i$, $i=0,1,2$ denote respectively the $i$-th derivatives of $r$ and $z$ with respect to $Z=R-R_m$ at $Z=0$, $R_m$ is the radius of the undeformed middle surface and the prime signifies differentiation with respect to $X$. It is clear from the above explicit expressions that $(p^{(1)},\bm{x}^{(2)})$ can be expressed in terms of $(p^{(0)},\bm{x}^{(0)},\bm{x}^{(1)})$ and $(p^{(0)},\bm{x}^{(1)})$ can be expressed in terms of $\bm{x}^{(0)}$.

In the case that the shell is a cylindrical tube in the reference configuration and is subjected to an axisymmetric motion described by $\bm{x}(\bm{X}, t)=r(R,X,t)\bm{e}_R+z(R,X,t)\bm{e}_{X}$, the refined shell equations \eqref{eq:ffinal12} and \eqref{eq:ffinal3} in cylindrical polar coordinates $(R,\Theta,X)$ take the form
\begin{align}
&\frac{\partial {S}^{(0)}_{XX}}{\partial X}=(1+\frac{1}{3}Kh^2)\rho \ddot{z}_0-\overline{q}_X,\\
\begin{split}
&\frac{\partial S^{(0)}_{XR}}{\partial X}-\frac{\partial S^{(0)}_{RX}}{\partial X}-\frac{S^{(0)}_{\Theta\Theta}}{R_m}+\frac{1}{3}h^2\frac{\partial^2 S^{(1)}_{XX}}{\partial X^2}\\
=&(1+\frac{1}{3}Kh^2)\rho\ddot{r}_0-\overline{q}_R+\frac{1}{3}h^2 (\rho  \frac{\partial \ddot{z}_1}{\partial X}-\frac{\partial \ddot{q}^{(1)}_{bX}}{\partial X})-\frac{\partial m_X}{\partial X},
\end{split}
\end{align}
where $r_0, z_0$ and $z_1$ have the same meanings as before.

\section{Boundary conditions}\label{app:bc}

Let $\partial \Omega$ denote the boundary of the middle surface, which is divided into two parts: the displacement boundary $\partial\Omega_0$ subjected to prescribed displacement and the traction boundary $\partial\Omega_q$ subjected to applied traction.  Let $s$ denote the arc length variable of $\partial\Omega$, and let $\bm{\tau}$ and $\bm{\nu}$ be respectively the unit tangent and outward normal vectors to $\partial\Omega$ such that $(\bm{\tau},\bm{n},\bm{\nu})$ forms a right-handed triple.

According to the 2d shell virtual work principle in \cite{YFD}, on the displacement edge $\partial\Omega_0$, we may impose
\begin{align}\label{eq:ub}
\bm{u}_{mt}=\hat{\bm{u}}_{mt},\quad u_{m3}=\hat{u}_{m3},\quad \alpha_{m}=\hat{\alpha}_{m}\iff \frac{u_{m3,\nu}}{1+\nabla\bm{u}_{mt}[\bm{\nu},\bm{\nu}]}=\tan(\hat{\alpha}_m),
\end{align}
where $\bm{u}=\bm{x}-\bm{X}$ is the displacement vector, $\bm{u}_m=\bm{u}^{(0)}$ and $\alpha_{m}=\arctan(u_{m3,\nu}/({1+\nabla\bm{u}_{mt}[\bm{\nu},\bm{\nu}]}))\pm n \pi$ are respectively the displacement and the rotation angle of the middle surface, and $\hat{\bm{u}}_m$ and $\hat{\alpha}_m$ are respectively the prescribed displacement and rotation angle of the same surface.  On the traction edge $\partial\Omega_q$, we may impose
\begin{align}
&2h \bm{S}^{(0)T}_{t}\bm{\nu}=\hat{\bm{q}}_t,\label{eq:work12}\\
\begin{split}
&2h\big((\bm{1}\bm{S}^{(0)}\bm{n}-\bm{1}\bm{S}^{(0)T}\bm{n})\cdot\bm{\nu}+\frac{1}{3}h^2(\bm{1}\nabla\cdot\bm{S}^{(1)}_t-\rho\ddot{\bm{x}}^{(1)}_t+\bm{q}_{bt}^{(1)})\cdot\bm{\nu}\\
&+\frac{1}{3}h^2(\bm{S}^{(1)T}_t[\bm{\nu},\bm{\tau}])_{,s}+\bm{m}_t\cdot\bm{\nu}\big)=\hat{q}_3,\label{eq:work3}
\end{split}\\
&2h(\frac{1}{3}h^2\bm{S}^{(1)T}_t[\bm{\nu},\bm{\nu}]+\frac{1}{3}h^2\bm{S}^{(1)T}[\bm{\nu},\bm{\tau}\times\bm{u}^{(1)}]+\frac{1}{6}h^2\bm{S}^{(0)T}[\bm{\nu},\bm{\tau}\times\bm{u}^{(2)}])=\hat{m}_3,\label{eq:workbending}
\end{align}
where  $\hat{\bm{q}}_t$ and $\hat{q}_3$ are respectively the applied in-plane and (total effective) shear forces (per unit arc length of $\partial\Omega_q$), and $\hat{m}_3$ is the applied bending moment about the middle surface.

\section{Expressions of the elastic moduli} \label{app:moduli}

For the strain energy function \eqref{eq:SE},  the associated first-order and second-order elastic moduli $\mathcal{A}^1$ and $\mathcal{A}^2$ are given by
\begin{align}
\begin{split}\label{eq:A1}
&\mathcal{A}^{1}(\bm{F})[\bm{A}]=c\bm{A}^T  + \sum_{i=4,6} \big(2k_1(I-1)\exp(k_2(I-1)^2)\bm{M}_i\otimes\bm{A}\bm{M}_i\\
&\qquad\qquad\quad\ +4k_1(1+2k_2(I-1)^2)\exp(k_2(I-1)^2)(\bm{A}\bm{M}_i\cdot\bm{F}\bm{M}_i)\bm{M}_i\otimes\bm{F}\bm{M}_i\big),
\end{split}\\
\begin{split}
&\mathcal{A}^2(\bm{F})[\bm{A},\bm{B}]
=\sum_{i=4,6}\big(4k_1(1+2k_2(I-1)^2)\exp(k_2(I-1)^2)((\bm{A}\bm{M}_i\cdot\bm{B}\bm{M}_i)\bm{M}_i\otimes\bm{F}\bm{M}_i\\
&\qquad\qquad\qquad\quad +(\bm{A}\bm{M}_i\cdot\bm{F}\bm{M}_i)\bm{M}_i\otimes\bm{B}\bm{M}_i+(\bm{B}\bm{M}_i\cdot\bm{F}\bm{M}_i)\bm{M}_i\otimes\bm{A}\bm{M}_i)\\
&\qquad\qquad\qquad\quad+16k_1k_2 (I-1)(3+2k_2(I-1)^2)\exp(k_2(I-1)^2)\\
&\qquad\qquad\qquad\quad \times (\bm{A}\bm{M}_i\cdot\bm{F}\bm{M}_i)(\bm{B}\bm{M}_i\cdot\bm{F}\bm{M}_i)\bm{M}_i\otimes\bm{F}\bm{M}_i\big)
\end{split}
\end{align}
with $\bm{M}_4=\bm{M}$ and $\bm{M}_6=\bm{M}'$. The components of $\mathcal{A}^1$ and $\mathcal{A}^2$ can be extracted by setting $\bm{A}=\bm{e}_i\otimes \bm{e}_j$ and $\bm{B}=\bm{e}_k\otimes \bm{e}_l$ with $(\bm{e}_1,\bm{e}_2,\bm{e}_3)$ being the standard basis vectors of $\mathbb{R}^3$.

\section{$O(h^2)$-correction to the incremental recurrence relations}\label{app:corr}
The next-order solution to \eqref{eq:ip0} and \eqref{eq:ix1} is given by
\begin{align}
&\delta p^{(0)b}=\frac{\bm{g}\cdot\bm{B}^{-1}\delta\bm{e}}{\bm{g}\cdot\bm{B}^{-1}\bm{g}},\quad \delta\bm{x}^{(1)b}=\bm{B}^{-1}(\delta{p}^{(0)b}\bm{g}-\delta\bm{e})
\end{align}
with $\delta\bm{e}$ defined by
\begin{align}\label{eq:y1}
\begin{split}
\delta\bm{e}&=p^{(0)b}\bm{F}^{(0)a-T}\delta\bm{F}^{(0)aT}\bm{F}^{(0)a-T}\bm{n}+K\delta\bm{S}^{(0)aT}\bm{n}-2H(\rho\delta\ddot{\bm{x}}^{(0)a}-\delta\bm{q}^{(0)}_b-\nabla\cdot\delta\bm{S}^{(0)a})\\
&\quad-\frac{1}{2}(\bm{\kappa}\bm{g}^\alpha)\cdot \delta\bm{S}^{(0)a}_{,\alpha}+\frac{1}{2}(\rho \delta\ddot{\bm{x}}^{(1)a}-\delta\bm{q}^{(1)}_b-\nabla\cdot\delta\bm{S}^{(1)a}).
\end{split}
\end{align}

\section{Expressions of the coefficient matrices $M(r,n,k,\omega)$ and $N(r,n,k,\omega)$}\label{app:coeff}

Let $\B$ denote the tensor of the first-order instantaneous modulus, which is related to the modulus $\A^1$ defined in \eqref{eq:A1} by the formula
\begin{align}
\B[\bm{A}]=\bm{F}\mathcal{A}^1[\bm{A}\bm{F}]\iff \B_{ijkl}=F_{im}F_{kn}\mathcal{A}^1_{mjnl}.
\end{align}
Note that the above expressions for $\B_{ijkl}$ are different from those given in  \cite{Ogden} (Problem 6.18) since the current material model is anisotropic.

Let us identify the indices $1,2,3$ with the $\theta$-, $z$- and $r$-directions respectively, and denote by $\lambda_1$, $\lambda_2$ and $\lambda_3$ the principal stretches in these directions. Then the coefficients $a_{2j}$, $a_{4j}$ and $a_{6j}$, $1\leq j\leq 6$ in \eqref{eq:exARRA} are given by
\begin{align}
\begin{split}
&a_{24}=a_{42}=a_{66}=0,\\
&r^2 \B_{3131} a_{21}=-  r^2 \rho\omega^2+\hat{k}^2 r^2\B_{2121} +r(\B_{1331}'+\overline{p}_r)+\B_{1313} +n^2(\B_{1111}-\B_{1133}-\B_{3113}),\\
&r \B_{3131} a_{22}=-r\B_{3131}'-\B_{3131},\\
&r \B_{3131} a_{23}=\hat{k} n (\B_{1122}+\B_{1221}-\B_{1133}-\B_{1331}),\\
& r^2 \B_{3131} a_{25}=n(r(\B_{1331}'+\overline{p}_r)+\B_{1111}+\B_{1313}-\B_{1133}-\B_{1331}),\\
&r \B_{3131} a_{26}=-n,\\
&r\B_{3232} a_{41}= \hat{k}n(\B_{1122}+\B_{1221}-\B_{2233}-\B_{2332}),\\
&  r^2 \B_{3232} a_{43}=-r^2 \rho\omega^2+\hat{k}^2r^2(\B_{2222}-\B_{2233}-\B_{2332})+n^2\B_{1212},\\
& r \B_{3232}a_{44}=-r\B_{3232}'-\B_{3232},\\
& r\B_{3232}a_{45}=\hat{k}(r(\B_{2332}'+\overline{p}_r)+\B_{1122}-\B_{2233}),\\
&\B_{3232}a_{46}=-\hat{k},\\
& r^2 a_{61}=n(r(\B_{1133}'-\B_{3333}'-\overline{p}_r)+\B_{3333}-\B_{1111}-\B_{1313}),\\
& r a_{62}= n(\B_{1133}+\B_{1331}-\B_{3333}),\quad r a_{63}=\hat{k}(r(\B_{2233}'-\B_{3333}'-\overline{p}_r)+\B_{2233}-\B_{1122}),\\
& a_{64}= \hat{k}(\B_{2233}+\B_{2332}-\B_{3333}),\\
&r^2 a_{65}= r^2\rho\omega^2-\hat{k}^2 r^2\B_{2323}+r(\B_{1133}'-\B_{3333}'-\overline{p}_r)+\B_{3333}-\B_{1111}-n^2\B_{1313}.
\end{split}
\end{align}
The coefficients $b_{3j}$, $1\leq j\leq 6$ in \eqref{eq:exARRA} are given by
\begin{align}
\begin{split}
&r b_{31}=n(\B_{1133}-\B_{3333}-\sigma_{3}), \quad  b_{32}=0,\quad b_{33}=\hat{k}(\B_{2233}-\B_{3333}-\sigma_{3}),\\
& b_{34}=0,\quad r b_{35}= \B_{1133}-\B_{3333}-\sigma_3,\quad  b_{36}=-1.
\end{split}
\end{align}
In the above expressions, the prime means $d/dr$, $\sigma_{3}=\lambda_3 \partial W/\partial \lambda_3$, and $\overline{p}_r=\sigma_3'+(\sigma_{3}-\sigma_{1})/r$ with $\sigma_1$ being $\sigma_1=\lambda_1\partial W/\partial \lambda_1$.


\end{document}